\documentclass[preprint,12pt]{elsarticle}

\usepackage{amssymb}

\usepackage{lineno}

\usepackage{hyperref}
\usepackage{listings}
\usepackage{xcolor} 
\usepackage{subcaption}

\usepackage{marginnote}
\usepackage[marginparwidth=2.5cm]{geometry}

\usepackage{multirow}
\usepackage{array}
\usepackage{enumitem}

\definecolor{gray}{rgb}{0.4,0.4,0.4}
\definecolor{darkblue}{rgb}{0.0,0.0,0.6}
\definecolor{cyan}{rgb}{0.0,0.6,0.6}

\lstdefinelanguage{XML}
{
  morestring=[b]",
  morestring=[s]{>}{<},
  morecomment=[s]{<?}{?>},
  keywordstyle=\color{cyan},
  moredelim=[s][\bfseries\color{darkblue}]{<}{\ },
  moredelim=[s][\bfseries\color{darkblue}]{</}{>},
  moredelim=[s][\color{black}]{(}{)},
  moredelim=[l][\bfseries\color{darkblue}]{/>},
  moredelim=[l][\bfseries\color{darkblue}]{>},
  morecomment=[s]{<?}{?>},
  morecomment=[s]{<!--}{-->},
  stringstyle=\color{black},
  identifierstyle=\color{cyan}
}

\lstdefinelanguage{SimplifiedXML}
{
    morestring=[b]",
    morecomment=[s]{<?}{?>},
    keywordstyle=\color{cyan},
    moredelim=[s][\bfseries\color{darkblue}]{>}{\ },
    moredelim=[s][\bfseries\color{darkblue}]{<}{\ },
    moredelim=[s][\bfseries\color{darkblue}]{</}{>},
    moredelim=[l][\bfseries\color{darkblue}]{/>},
    morecomment=[s]{<?}{?>},
    morecomment=[s]{<!--}{-->},
    identifierstyle=\color{black},
    stringstyle=\color{black},
    morekeywords={processRef, sourceRef, targetRef}
}

\lstdefinelanguage{JSON}
{
    morestring=[b]",
    morecomment=[s]{<?}{?>},
    keywordstyle=\color{cyan},
    morecomment=[s]{<?}{?>},
    morecomment=[s]{<!--}{-->},
    identifierstyle=\color{black},
    stringstyle=\color{black},
    morekeywords={\$type, \$parent, id, name, processRef, lanes, flowNodeRef, sourceRef, targetRef},
    morecomment=[s]{<?}{?>},
    morecomment=[s]{<!--}{-->},
    stringstyle=\color{black},
}

\journal{Decision Support Systems}

\usepackage{ifthen}
\newboolean{forreview}
\setboolean{forreview}{false} 
\ifthenelse{\boolean{forreview}}
{\newcommand{\ifReview}[2]{{#1}}}
{\newcommand{\ifReview}[2]{{#2}}}

\begin{document}

\begin{frontmatter}



\title{Leveraging Large Language Models for Enhanced Process Model Comprehension}


\ifReview{}{
\author[inst1,inst2]{Humam Kourani} 
\author[inst1,inst2]{Alessandro Berti}
\author[inst1]{Jasmin Hennrich}
\author[inst1]{Wolfgang Kratsch}
\author[inst1]{Robin Weidlich}
\author[inst1,inst2]{Chiao-Yun Li} 
\author[inst1,inst2]{Ahmad Arslan} 
\author[inst1,inst2]{Wil M. P. van der Aalst}
\author[inst1,inst2]{Daniel Schuster} 

 \affiliation[inst1]{organization={Fraunhofer Institute for Applied Information Technology FIT},
     addressline={Schloss Birlinghoven}, 
             city={Sankt Augustin},
             postcode={53757}, 
             country={Germany}}

 \affiliation[inst2]{organization={RWTH Aachen University},
             addressline={Ahornstraße 55}, 
             city={Aachen},
             postcode={52074}, 
             country={Germany}}
}

\begin{abstract}

In Business Process Management (BPM), effectively comprehending process models is crucial yet poses significant challenges, particularly as organizations scale and processes become more complex. This paper introduces a novel framework utilizing the advanced capabilities of Large Language Models (LLMs) to enhance the interpretability of complex process models. We present different methods for abstracting business process models into a format accessible to LLMs, and we implement advanced prompting strategies specifically designed to optimize LLM performance within our framework. 
Additionally, we present a tool, AIPA, that implements our proposed framework and allows for conversational process querying. We evaluate our framework and tool by i) an automatic evaluation comparing different LLMs, model abstractions, and prompting strategies and ii) a user study designed to assess AIPA's effectiveness comprehensively.
Results demonstrate our framework's ability to improve the accessibility and interpretability of process models, pioneering new pathways for integrating AI technologies into the BPM field. 
\end{abstract}

\begin{graphicalabstract}
\includegraphics[width=\textwidth]{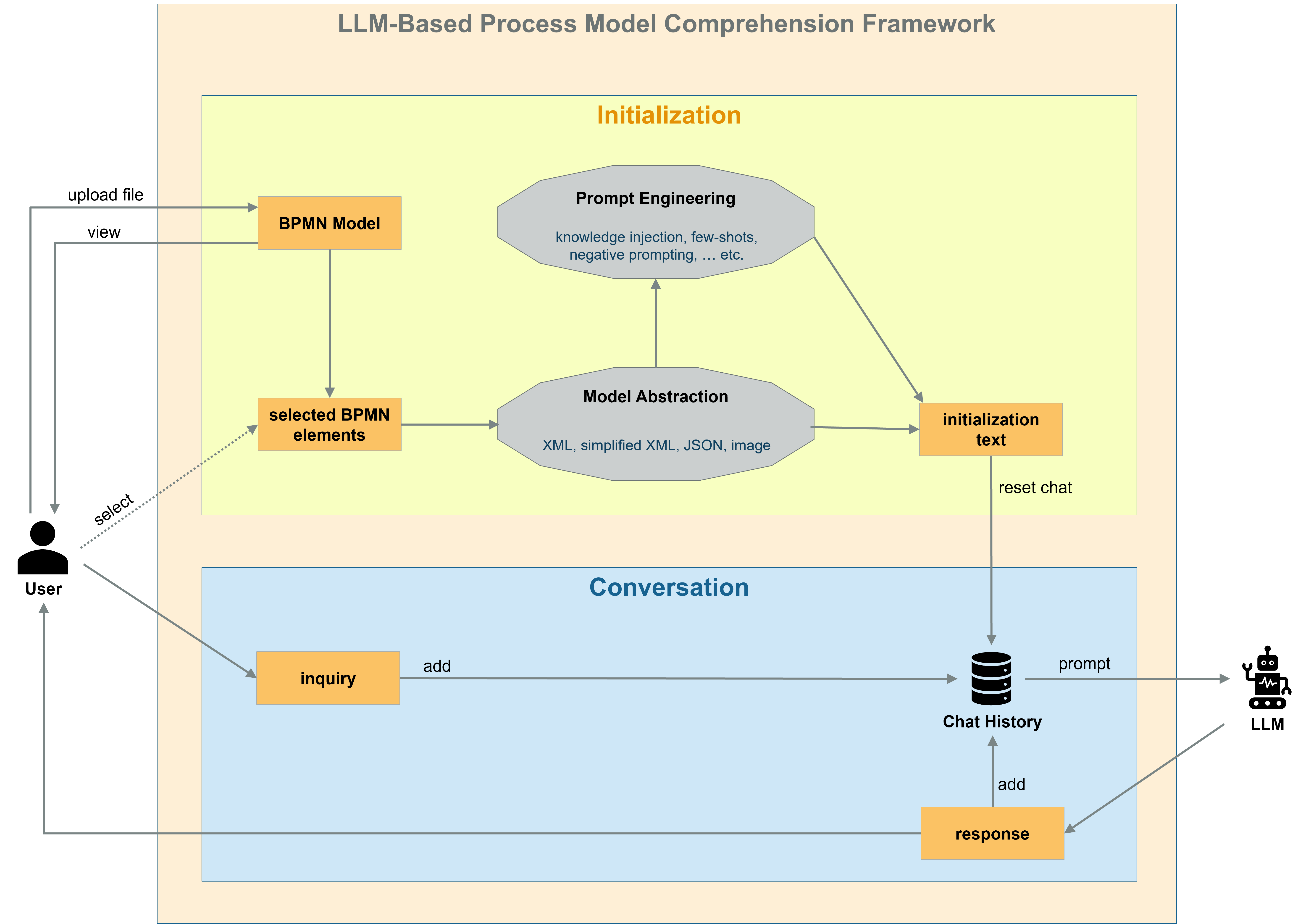}
\end{graphicalabstract}

\begin{highlights}
\item Cutting-edge LLMs can effectively interpret abstractions of business process models.
\item Prompting techniques enhance the LLMs' comprehension of process models.
\item Advanced LLMs can understand not only textual abstractions but also images. 
\item Recent open-source LLMs demonstrate comparable effectiveness to commercial solutions.
\item A case study shows promising results with suggestions for more concise responses.

\end{highlights}

\begin{keyword}
Process Model Comprehension \sep Business Process Management \sep Large Language Models \sep Generative AI
\end{keyword}

\end{frontmatter}


\section{Introduction}\label{sec:intro}

    \emph{Business Process Management (BPM)} represents a management approach focusing on aligning an organization's processes with its strategic objectives. This includes process documentation, automation, integration, and continuous process improvement. Using BPM allows organizations to optimize performance, manage change, and achieve operational excellence \cite{DBLP:books/sp/DumasRMR18}.

    In the context of BPM, \emph{process models} serve as the primary artifact for depicting the flow of activities, data, events, and organizational units in a process. Process models facilitate the analysis, simulation, optimization, and automation of business processes. In today’s competitive market, high-quality process models are pivotal for enterprises seeking to enhance their operational efficiency and the quality of their products and services. However, the inherent complexity of real-world business processes often results in intricate models that can be difficult to understand and manage. This complexity can lead to higher costs and more errors during the maintenance and improvement of the processes.

    The Business Process Model and Notation (BPMN) \cite{DBLP:books/el/15/RosingWCM15} has become the de facto standard for modeling business processes and is widely used in industry. While BPMN offers a diverse range of elements and constructs, typical usage in industry centers around a limited, commonly used subset \cite{DBLP:conf/caise/MuehlenR08,DBLP:books/daglib/p/MuehlenR13}. However, the availability of additional, more advanced elements allows for the modeling of specialized or complex scenarios. This capability enhances BPMN's versatility but also complicate model comprehension, increasing the risk of cognitive overload for users \cite{DBLP:journals/bpmj/Recker10,DBLP:journals/jcis/Bera12}. This complexity acts as a barrier, hindering users from fully comprehending BPMN models.


    In addressing the challenge of comprehending complex process models, leveraging new technologies in AI holds promise. Among these technologies, \emph{Large Language Models (LLMs)} stand out for their advanced capabilities in natural language processing and pattern recognition. Advanced LLMs, such as \emph{gpt-4} \cite{DBLP:journals/corr/abs-2303-08774}, are trained on vast amounts of text data, enabling them to generate human-like text and engage in natural language conversations \cite{DBLP:journals/corr/abs-2303-18223}. Due to their comprehensive training data, LLMs contain a wealth of process-related domain knowledge that can facilitate process model comprehension \cite{DBLP:conf/aaai/GuZYZWZJXLWHXHL24}. Moreover, LLMs possess reasoning capabilities, allowing them to recognize, analyze, and make inferences from textual data in various contexts \cite{DBLP:conf/acl/0009C23}. These reasoning capabilities might enable LLMs to comprehend process models, identify relationships between elements, and interpret process structures accurately.

     
    %

   
In this paper, we present a novel framework that leverages the capabilities of LLMs to enhance the interpretability and accessibility of complex BPMN models. By abstracting process models into different formats and employing advanced prompt engineering techniques, we guide LLMs to comprehend the different process structures and relationships. Our framework enables users to interact with the LLMs, gaining deep insights into complex processes without requiring technical expertise in modeling languages. Furthermore, we present our tool, AIPA (AI-Powered Process Analyst), which integrates our framework with state-of-the-art LLMs. Our research is structured around the following key research questions:
    \begin{itemize}
        \item RQ1: How does the choice of abstraction method for BPMN models affect their comprehension by LLMs?
        \item RQ2: What prompting strategies can enhance the comprehension capabilities of LLMs for BPMN models?
        \item RQ3: How effectively can state-of-the-art LLMs understand and interpret BPMNs models?
        \item RQ4: How does the application of LLMs influence user comprehension of BPMN models in practical scenarios?
    \end{itemize}

    The remainder of the paper is structured as follows. \autoref{sec:rel} discusses related work. In \autoref{sec:framework}, we introduce our framework for LLM-based process model comprehension. In \autoref{sec:tool}, we present our tool, AIPA. We evaluate our framework on different textual abstractions and prompting strategies in \autoref{sec:ev}, and we conduct a user study to assess the effectiveness and usability of AIPA in \autoref{sec:study}. \autoref{sec:future} discusses the limitations of our framework and proposes ideas for future work. Finally, \autoref{sec:conc} concludes this paper.

\section{Related Work}\label{sec:rel}


The complexity of comprehending business process models has been extensively explored. In \cite{zhou_business_2023}, the authors conducted a comprehensive review of metrics for assessing business process complexity, identifying different dimensions of complexity. In \cite{DBLP:journals/softx/AndaloussiLW23}, the authors empirically evaluated the cognitive difficulty associated with comprehending process models, focusing on specific process constructs. Their findings revealed that repetition and exclusive choices impose a higher cognitive load compared to concurrent and sequential tasks. The study \cite{winter2023comparative} assessed the comprehension of BPMN models by healthcare associates, highlighting the challenges they face in understanding complex BPMN models. These studies underscore the need for new methods to enhance the interpretability and accessibility of process models.



\emph{Process model querying} methods \cite{DBLP:reference/bdt/Polyvyanyy19} play a crucial role in enhancing the accessibility and usability of business process models. These methods can be categorized into two main categories: \emph{model-specific querying} \cite{DBLP:books/sp/22/DelfmannRHCD22,DBLP:books/sp/22/StorrleA22,DBLP:books/sp/22/FrancescomarinoT22,DBLP:books/sp/22/KammererPR22} and \emph{models repository querying} \cite{DBLP:books/sp/22/Polyvyanyy22a,DBLP:books/sp/22/ProiettiTS22} (i.e., finding the models in a repository satisfying some given constraint). All the proposed approaches allow for a large number of queries.
However, the model-specific approaches are limited by i) lack of domain knowledge on the underlying process;
ii) prototypal implementation;
iii) the user needs to learn a new querying language (either graphical or textual).
These limitations highlight the need for more intuitive and accessible querying methods, which our framework addresses by leveraging LLMs to facilitate conversational process querying.


AI systems can help non-expert analysts interpret BPMN models by automatically analyzing, extracting, and summarizing relevant information from process models \cite{DBLP:journals/eswa/RosaRADMDG11}. In \cite{DBLP:conf/abict/LigezaP12}, the authors present a logical model for analyzing BPMN models using AI techniques. Their approach demonstrates the ability of AI systems to detect patterns, errors, and inconsistencies, thereby enhancing the analysis of BPMN models. In \cite{DBLP:conf/rcis/CascianiBCM24}, the authors leverage conversational AI to create interactive systems that can understand and respond to natural language inputs, thus facilitating more intuitive and efficient management of business processes. These advancements underscore the potential of AI technologies to simplify the interpretation of complex process models.

Recently, some LLM-based model querying methods have been proposed. In \cite{DBLP:conf/bpm/Berti0A23}, Petri nets are textually abstracted for process description and improvement purposes. Declarative models involving the control flow and time perspectives are texutally abstracted in \cite{DBLP:journals/corr/abs-2307-12701}, allowing for a range of queries. In \cite{bernardi2024conversing}, a double-faced approach involving fine-tuning open-source LLMs and retrieval augmented generation techniques on a specific BPMN process model is proposed. However, the results of the assessment are not convincing, achieving low percentages on simple tasks such as tasks and flow existence. The approach described in \cite{fahland2024well} aims to explain decision points in a process exploiting both the structure of the process model and information extracted from the event log (for instance, causal relations and Shapley values). 
A limitation stated in the paper is that the implemented tool does not allow for iterative cycles, being limited to the initial response. Moreover, it is not possible to express questions unrelated to decision points.

\section{LLM-Based Process Model Comprehension Framework}\label{sec:framework}
This section introduces our framework for LLM-based process model comprehension. We outline the high-level architecture of the framework, and we propose several methods for abstracting process models and various prompting strategies aimed at optimizing LLM performance. 

    \subsection{Architecture} \label{sec:arch}

    \begin{figure}[!t]
    \centering        
    \includegraphics[width=0.9\textwidth]{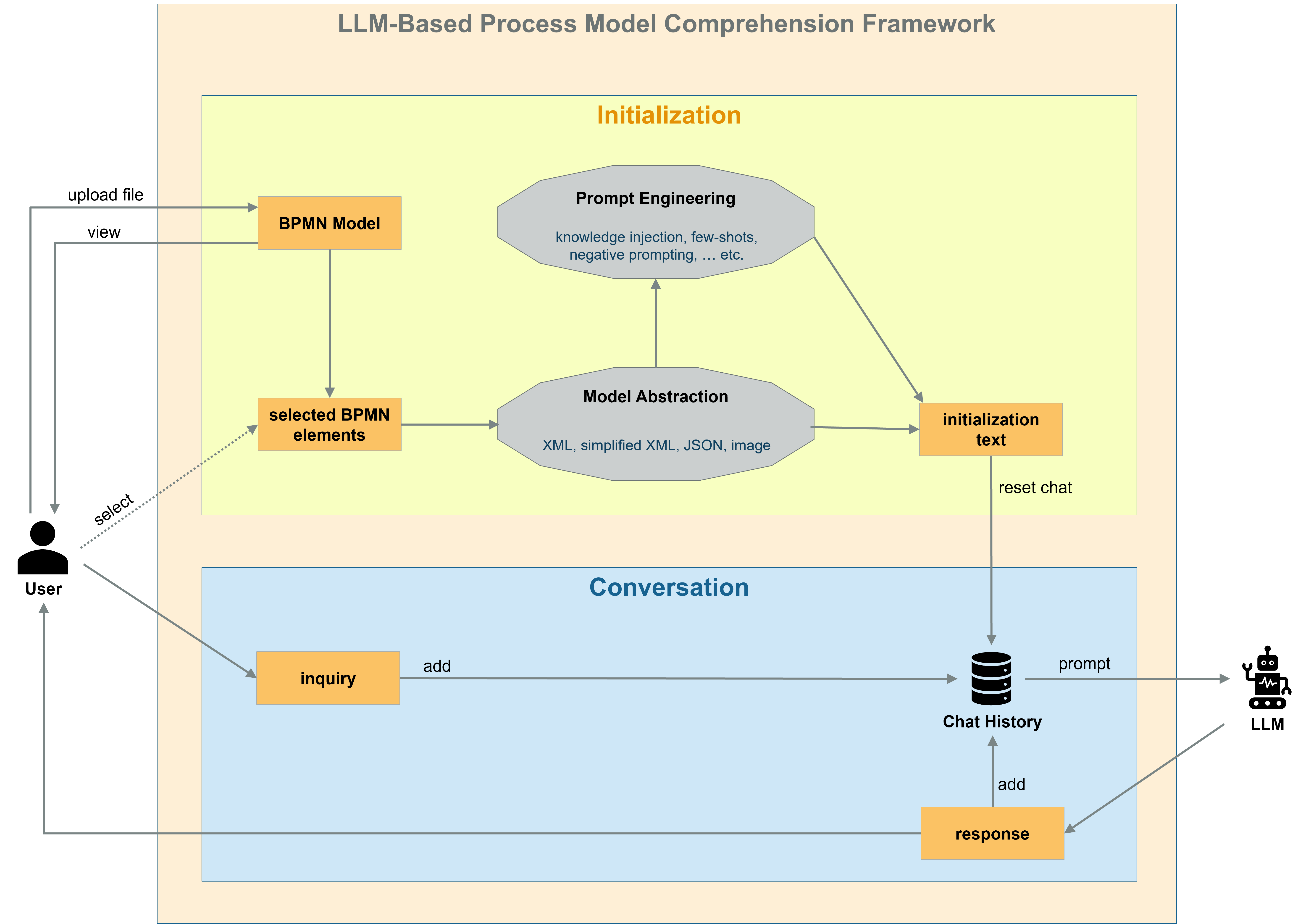}
    \caption{LLM-based process model comprehension framework.}
     \label{fig:overview}
 \end{figure}
        
         \autoref{fig:overview} illustrates the architecture of our LLM-based process model comprehension framework. 
         The process initiates with a user uploading a BPMN model, which is automatically abstracted into an input format that an LLM can process (cf. \autoref{sec:abst}). For targeted analysis, users can optionally select specific parts of the BPMN model. Only these selected elements are then included in the generated abstraction to facilitate the analysis of specified areas of interest.
         
          To aid the LLM in effectively understanding and analyzing the BPMN model, the model abstraction is augmented with tailored instructions. These incorporate various prompting strategies designed to optimize the LLM's response (cf. \autoref{sec:prompt}). 
          
          The user then poses an inquiry about the BPMN model, which is integrated into a prompt that combines the textual representation, the applied prompting strategies, and the user’s inquiry. This enriched prompt is submitted to the LLM, which processes the input and generates a response. The response is then forwarded to the user. The interaction is dynamic, allowing the user to pose follow-up questions. Each follow-up question is integrated into a new prompt, maintaining the conversation history, and submitted to the LLM, which then generates subsequent responses.

\subsection{BPMN Abstraction}\label{sec:abst}
The abstraction of BPMN models is a crucial component of our framework, enabling their transformation into formats that can be effectively processed by LLMs. This section details the four abstraction formats supported by our framework: XML, simplified XML, JSON, and image.

\paragraph{XML}
BPMN Models of the  2.0 standard are typically stored in an XML format \cite{DBLP:books/el/15/RosingWCM15}. This comprehensive format encapsulates all visual and structural aspects of the process, including tasks, events, gateways, and detailed attributes such as positions, styles, and additional metadata. The Full XML abstraction retains all these details, conforming with BPMN 2.0 standard. \autoref{lst:fullXML} shows the full XML abstraction of an example BPMN model.



\begin{figure*}
\begin{lstlisting}[caption={Full XML abstraction for an example BPMN model.}, frame=single, label={lst:fullXML},  basicstyle=\scriptsize\ttfamily, language=XML]
<?xml version="1.0" encoding="UTF-8"?>
<bpmn:definitions
    xmlns:bpmn="http://www.omg.org/spec/BPMN/20100524/MODEL" 
    xmlns:bpmndi="http://www.omg.org/spec/BPMN/20100524/DI"
    xmlns:di="http://www.omg.org/spec/DD/20100524/DI"
    xmlns:dc="http://www.omg.org/spec/DD/20100524/DC"
    xmlns:xsi="http://www.w3.org/2001/XMLSchema-instance" 
    id="Definitions_1" targetNamespace="http://bpmn.io/schema/bpmn" 
    exporter="bpmn-js (https://demo.bpmn.io)" exporterVersion="17.6.4">
  <bpmn:collaboration id="col_1">
    <bpmn:participant id="par_1" name="My Process" processRef="pro_1"/>
  </bpmn:collaboration>
  <bpmn:process id="pro_1" isExecutable="false">
    <bpmn:laneSet>
      <bpmn:lane id="lane_1" name="My Resource">
        <bpmn:flowNodeRef>task_1</bpmn:flowNodeRef>
        <bpmn:flowNodeRef>event_1</bpmn:flowNodeRef>
      </bpmn:lane>
    </bpmn:laneSet>
    <bpmn:task id="task_1" name="Task 1">
      <bpmn:incoming>flow_1</bpmn:incoming>
    </bpmn:task>
    <bpmn:startEvent id="event_1" name="Start">
      <bpmn:outgoing>flow_1</bpmn:outgoing>
    </bpmn:startEvent>
    <bpmn:sequenceFlow id="flow_1" sourceRef="event_1" targetRef="task_1"/>
  </bpmn:process>
  <bpmndi:BPMNDiagram id="BPMNDiagram_1">
    <bpmndi:BPMNPlane id="BPMNPlane_1" bpmnElement="col_1">
      <bpmndi:BPMNShape id="di_1" bpmnElement="par_1" isHorizontal="true">
        <dc:Bounds x="152" y="80" width="498" height="190"/>
        <bpmndi:BPMNLabel/>
      </bpmndi:BPMNShape>
      <bpmndi:BPMNShape id="di_2" bpmnElement="lane_1" isHorizontal="true">
        <dc:Bounds x="182" y="80" width="468" height="190"/>
        <bpmndi:BPMNLabel/>
      </bpmndi:BPMNShape>
      <bpmndi:BPMNShape id="di_3" bpmnElement="task_1">
        <dc:Bounds x="470" y="134" width="100" height="80"/>
        <bpmndi:BPMNLabel/>
      </bpmndi:BPMNShape>
      <bpmndi:BPMNShape id="_BPMNShape_StartEvent_2" bpmnElement="event_1">
        <dc:Bounds x="302" y="156" width="36" height="36"/>
        <bpmndi:BPMNLabel>
          <dc:Bounds x="308" y="192" width="25" height="14"/>
        </bpmndi:BPMNLabel>
      </bpmndi:BPMNShape>
      <bpmndi:BPMNEdge id="flow_1_di" bpmnElement="flow_1">
        <di:waypoint x="338" y="174"/>
        <di:waypoint x="470" y="174"/>
      </bpmndi:BPMNEdge>
    </bpmndi:BPMNPlane>
  </bpmndi:BPMNDiagram>
</bpmn:definitions>
\end{lstlisting}
\end{figure*}

\paragraph{Simplified XML (SXML)} 
We designed this abstraction format to reduce the complexity of the standard XML format by deliberately omitting non-essential elements such as layout information, styling, and metadata. These elements are considered irrelevant to the core logical structure of the BPMN model. SXML retains the original XML's hierarchical structure but includes only the fundamental components such as swimlanes, tasks, gateways, and connections. This abstraction provides a concise representation that highlights the operational aspects of the BPMN model without the additional complexity of visual details. \autoref{lst:simpXML} shows the simplified XML abstraction of the same BPMN model from \autoref{lst:fullXML}.

\begin{figure*}[!t]
\begin{lstlisting}[caption={Simplified XML abstraction for an example BPMN model.}, frame=single, label={lst:simpXML},  basicstyle=\scriptsize\ttfamily, language=SimplifiedXML]
<definitions Definitions_1>
    <collaboration col_1>
        <participant par_1> (My Process)
          - processRef: pro_1
        </participant>
    </collaboration>
    <process pro_1>
        <laneSet>
            <lane lane_1> (My Resource)
                <flowNodeRef (task_1)/>
                <flowNodeRef (event_1)/>
            </lane>
        </laneSet>
        <task task_1 (Task 1)/>
        <startEvent event_1 (Start)/>
        <sequenceFlow flow_1>
          - sourceRef: event_1
          - targetRef: task_1
        </sequenceFlow>
    </process>
</definitions>
\end{lstlisting}
\end{figure*}
    
\paragraph{JSON} 
We designed the JSON abstraction to restructure the BPMN data into an attribute-based representation. Unlike the hierarchical structure of XML, the JSON format organizes BPMN elements into a flat list, where each element is associated with a set of attributes that encapsulate all necessary information. Similarly to the Simplified XML abstraction, the JSON abstraction retains only the essential components and excludes non-essential attributes like styling and layout information. By doing so, it enables LLMs to process the structural and operational aspects of the process model efficiently, focusing on the attributes that are most relevant for analysis. \autoref{lst:json} shows the JSON abstraction of the same BPMN model from \autoref{lst:fullXML}.

\begin{figure*}[!t]
\begin{lstlisting}[caption={JSON abstraction for an example BPMN model.}, frame=single, label={lst:json},  basicstyle=\scriptsize\ttfamily, language=JSON]
- { $type: bpmn:Collaboration, id: col_1, $parent: Definitions_1 }
- { $type: bpmn:Participant, id: par_1, name: My Process, processRef: pro_1, $parent: col_1 }
- { $type: bpmn:Task, id: task_1, name: Task 1, lanes: (lane_1), $parent: pro_1 }
- { $type: bpmn:StartEvent, id: event_1, name: Start, lanes: (lane_1), $parent: pro_1 }
- { $type: bpmn:SequenceFlow, id: flow_1, sourceRef: event_1, targetRef: task_1, $parent: pro_1 }
- { $type: bpmn:Lane, id: lane_1, name: My Resource, flowNodeRef: (task_1, event_1) }
\end{lstlisting}
\end{figure*}
    
\paragraph{Image (PNG)}
Recent advancements in LLMs have extended their capabilities beyond text processing to include support for image inputs. This enhancement allows for the processing and analysis of non-textual data, offering a broader range of applications. We leverage these capabilities through the image abstraction of BPMN models. This abstraction involves transforming the graphical BPMN models into PNG images. This format captures the visual layout of the model, preserving the spatial arrangement of the different BPMN elements. By feeding the generated image into an advanced model, its image processing capabilities are used to interpret the BPMN model’s content and structure.

\subsection{Prompt Engineering Techniques}\label{sec:prompt}
    \emph{Prompt engineering} refers to the techniques used to instruct LLMs and guide them towards desired outputs. This includes providing additional information in the prompt to better inform the LLM about the requirements of the task and the domain knowledge required to perform it. The goal is to optimize the LLM's performance and improve the relevance and quality of its outputs. 
    
    In our framework, we support several prompting techniques that can be combined to optimize the understanding of BPMN models by LLMs. This section outlines these techniques, and the full prompts are available at \url{https://github.com/humam-kourani/AIPA/tree/main/evaluation}.

\subsubsection*{Role Prompting}
One problem of LLMs is their \emph{laziness}, which was empirically studied in \cite{DBLP:conf/acl/TangKH023}. Laziness should be interpreted as the tendency to reduce the effort to accomplish a task and can be either observed in generic answers, statements without explanations, or missing important parts of the original prompt. Some strategies have been proposed to help mitigate laziness. One strategy that can be used to migrate this issue is \emph{Role prompting} \cite{DBLP:journals/corr/abs-2308-07702}. This strategy involves configuring the prompt to position the LLM as a domain expert \cite{DBLP:journals/corr/abs-2305-14688}. As illustrated in \autoref{lst:role}, our framework includes two implementations of role prompting: 
\begin{itemize}

    \item \textit{Process Modeling Expert (S1):}
    We assign the LLM the role of an expert in business process modeling and the BPMN 2.0 standard.
    \item \textit{Process Owner (S2):}
    We assign the LLM the role of a process expert familiar with the domain of the provided process, instructing it to use its domain knowledge to fill in any missing gaps when analyzing the process.
\end{itemize}

\subsubsection*{Non-Technical Abstraction (S3)}
As shown in \autoref{lst:nontech}, we instruct the LLM to avoid technical terms and use natural language to describe the behavior of the underlying process. This technique ensures that explanations and analyses are accessible to users who may not be familiar with the specialized terminologies of the BPMN 2.0 standard.

\subsubsection*{Chain of Thoughts (S4)}
LLMs are subject to problems such as \emph{hallucinations} \cite{DBLP:conf/emnlp/JiYXLIF23} (the answer is partly unrelated to the original question) and \emph{non-determinism} \cite{DBLP:journals/corr/abs-2308-02828} (the answer changes in different sessions). The chain-of-thoughts technique aims to improve the interpretability and reasoning capabilities of the LLM by explicitly requesting the motivations and intermediate reasoning steps it employs to answer the question \cite{DBLP:conf/nips/Wei0SBIXCLZ22}. \autoref{lst:chain} illustrates our implementation of the chain of thoughts method.

\begin{figure*}[!t]
\begin{lstlisting}[caption={Implementation of role prompting.}, frame=single, label={lst:role},  basicstyle=\scriptsize\ttfamily]
- Your role: You are an expert in business process modeling and the BPMN 2.0 standard. I will give 
you a textual representation of a full BPMN model or only selected elements of a BPMN (e.g., a set of 
tasks and flows) and ask you questions about the process. You are supposed to answer the questions 
based on your understanding of the provided model.
- Please take the role of a process expert who is familiar with the domain of the provided process,
and use your domain knowledge to better understand and analyze the process, filling in any missing 
gaps.

\end{lstlisting}

\begin{lstlisting}[caption={Instructing the LLM to use non-technical language.}, frame=single, label={lst:nontech},  basicstyle=\scriptsize\ttfamily]
- Please answer in natural language, so that any user not familiar with the BPMN standard can 
understand your answer without any technical knowledge; i.e., avoid technical terms like Task, Gate,
flow, lane, etc.; rather use natural language to describe the behavior of the underlying process.
\end{lstlisting}

\begin{lstlisting}[caption={Implementation of chain of thoughts.}, frame=single, label={lst:chain},  basicstyle=\scriptsize\ttfamily]
- Where possible (especially for complex queries), please share the chain of thoughts or reasoning
behind your answers. This helps in understanding how you arrived at your conclusion.
\end{lstlisting}
\end{figure*}

\subsubsection*{Knowledge Injection (S5)}
LLMs are trained on a vast corpus of knowledge, but they might still miss context-specific information. To address this issue, \emph{fine-tuning} techniques \cite{DBLP:conf/icmla/KaziK23} revise the parameters of LLMs to include additional information about the task in question. However, due to the high computational costs associated with fine-tuning, an alternative is to engineer the prompts to include additional information. This strategy, known as \emph{knowledge injection} \cite{DBLP:conf/esws/MartinoIT23}, involves enriching prompts with task-specific information that the LLM may not have encountered during its initial training. 

The BPMN standard is inherently complex, encompassing a wide range of elements and constructs. To balance the comprehensiveness of the information with the constraints of the LLM's context limit, we provide an informative summary of BPMN essential elements. The injected knowledge covers the following BPMN elements:
\begin{itemize}
    \item Flow objects: events, gateways, tasks, sub-processes, transactions, and call activities.
    \item Connections: sequence flows, message flows, and associations.
    \item Further elements: pools, lanes, data objects, groups, and annotations.
\end{itemize}

\subsubsection*{Few-Shot Learning (S6)}
This method leverages the LLM's ability to learn from limited examples by providing several pairs of example inputs and their expected outputs \cite{DBLP:conf/nips/BrownMRSKDNSSAA20}. This choice over \emph{zero-shot}, \emph{one-shot}, and \emph{many-shot} learning approaches stems from our goal to optimize the balance between model training efficiency and output accuracy. Zero-shot learning, where the model generates answers without prior related examples, often leads to less accurate or generalized responses due to the absence of context-specific training. One-shot learning provides a single example, which may not sufficiently capture the complexity or variability of the task. Many-shot learning, though potentially more effective due to a broader range of examples, requires a substantially larger dataset.


Given the constraints of LLM prompt size and processing capacity, we elected to utilize only one BPMN model for training, complemented with five pairs of question-and-answer examples. While we anticipate that including a broader array of examples across more models might yield more robust capabilities in BPMN comprehension, the current setup was chosen to maintain a reasonable trade-off between the prompt size and the richness of the training data. This strategic decision allows us to experiment within a feasible framework while setting the stage for potentially scaling up the number of examples or integrating more diverse models in the future to enhance the LLM's performance. 

The BPMN model we use for implementing few-shot learning is the credit scoring process, available at \url{https://github.com/camunda/bpmn-for-research}. We crafted a set of five questions designed to challenge the LLM across various aspects of the process: starting the process, handling immediate outcomes, dealing with delays, and finalizing results. Moreover, the questions target different BPMN elements: tasks, events, pools, gateways, and message flows. The provided expected outputs for these questions were designed to not only provide accurate responses but also to use plain, accessible language that enhances comprehension for users without technical knowledge. \autoref{lst:few-shot} illustrates our few-shot learning example pairs.

\begin{figure*}[!t]
\begin{lstlisting}[caption={Examples used for few-shot learning. Lines that extend beyond the displayed text are abbreviated with ``...'' to keep it compact.}, frame=single, label={lst:few-shot},  basicstyle=\scriptsize\ttfamily]
- Let us consider the following textual representation of an example process: - { $type: ...
- These are example pairs of input and expected output:
Input 0: How does the bank start the credit scoring process?
Expected output 0: A bank clerk uses their software to request a credit score for a customer, ...
Input 1: What happens after the bank sends a scoring request to the agency?
Expected output 1: The agency performs an initial credit scoring, and if this initial scoring ...
Input 2: What occurs if the initial credit scoring doesn't give an immediate result?
Expected output 2: The agency informs the bank's system about the delay and starts a more ...
Input 3: How is the clerk informed of the final credit scoring result?
Expected output 3: Once the scoring is completed and the result is sent back to the bank's ...
Input 4: What does the bank do if the credit score is delayed?
Expected output 4: The bank's system sends a notification to the clerk to report the delay, ...
\end{lstlisting}

\begin{lstlisting}[caption={Extending the few-shot pairs by including examples of undesired outputs. Lines that extend beyond the displayed text are abbreviated with ``...'' to keep it compact.}, frame=single, label={lst:neg},  basicstyle=\scriptsize\ttfamily]
- Now, I will give you example bad outputs for the same example questions, so you try to avoid ...
Bad output for question 0: The bank initiates a BPMN Collaboration, invoking a Participant ...
Bad output for question 1: Upon receipt of the scoring request, an IntermediateCatchEvent is ...
Bad output for question 2: If the initial scoring doesn't resolve, a conditional ...
Bad output for question 3: The final scoring outcome, after processing through a sequence of ...
Bad output for question 4: In the event of a scoring delay, an IntermediateCatchEvent captures ...
\end{lstlisting}
\end{figure*}


\subsubsection*{Negative Prompting (S7)} Negative prompting involves guiding the LLM by explicitly stating what should be avoided in its responses \cite{DBLP:journals/corr/abs-2305-16807}. We enhance our few-shot learning demonstrations by including examples of undesired outputs as shown in \autoref{lst:neg}. We generated these responses by providing a textual abstraction of the BPMN model and the questions to GPT-3.5 without implementing any prompting strategies. The resulting answers include inaccuracies and misleading information. Additionally, they tend to be overly complex, filled with unnecessary technical details that obscure the clarity and relevance of the responses.

\section{Tool Support}\label{sec:tool}
In this section, we present our tool AIPA (AI-Powered Process Analyst) which integrates our LLM-based process model comprehension framework, as detailed in \autoref{sec:framework}, with OpenAI's LLMs. This integration is flexible, allowing users to configure the tool to use any OpenAI model. The tool can be downloaded from \url{https://github.com/humam-kourani/AIPA} and used following the instructions contained in the \textbf{README.md} file.

AIPA uses JSON as the default textual abstraction format to represent BPMN models. This choice was informed by our evaluation experiments in \autoref{sec:ev}, which show that the JSON and Simplified XML formats provided a clearer and more effective simplification compared to the other abstraction formats we proposed in \autoref{sec:abst}. The prompting strategies proposed in \autoref{sec:prompt} are all implemented in AIPA and enabled by default.

\begin{figure*}[!t]
\centering
\includegraphics[width=0.8\textwidth]{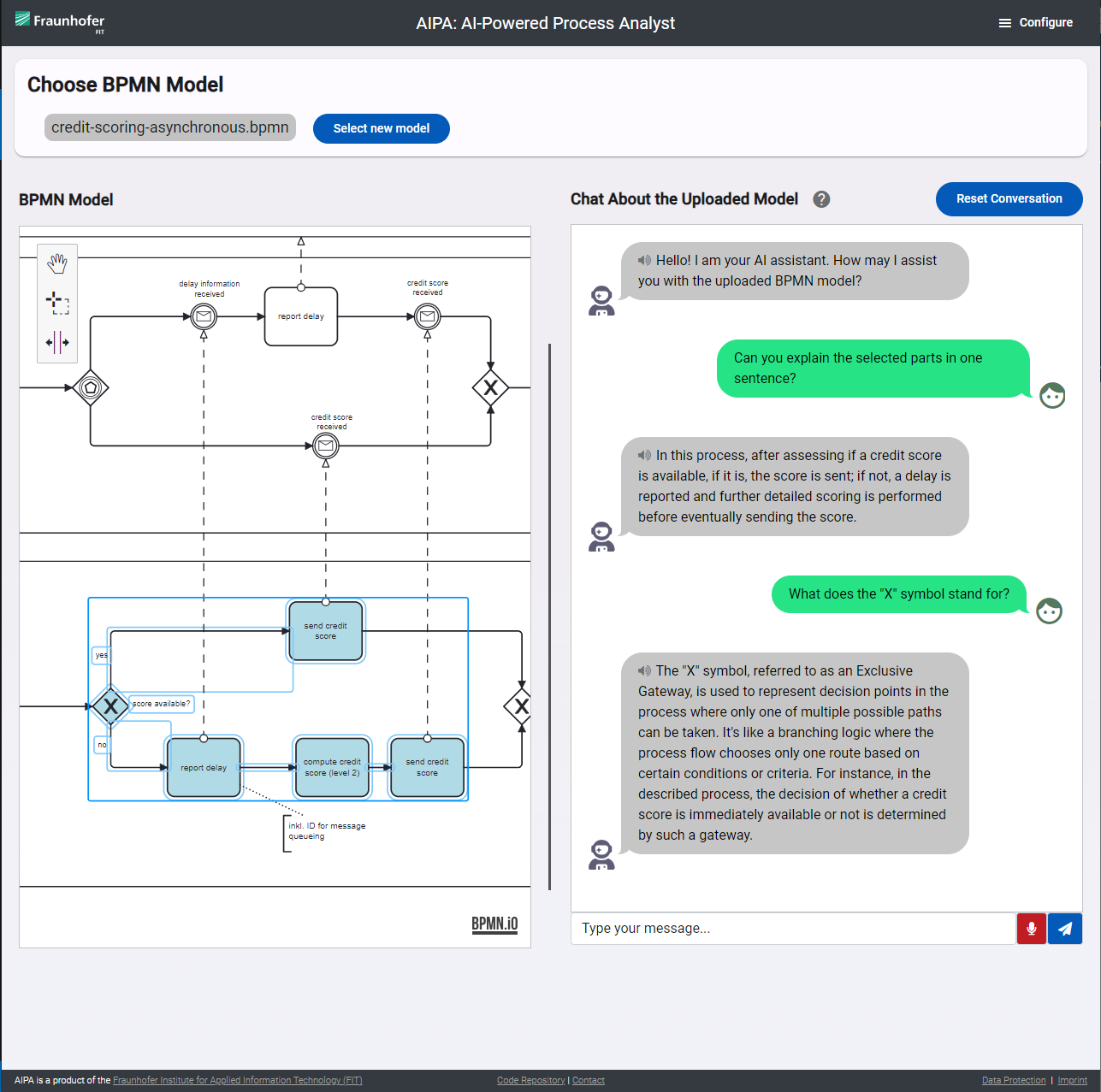}
\caption{Screenshot of AIPA.}
\label{fig:AIPA}
\end{figure*}

\autoref{fig:AIPA} shows a screenshot of AIPA. The user can upload a BPMN model and start a chat about the uploaded model with an AI assistant. When users select specific elements within a BPMN model, the tool generates a JSON representation containing only those selected elements. This focused abstraction ensures that the LLM processes only the relevant parts of the model, enhancing both the efficiency and relevance in the analysis. Additionally, our tool supports voice interactions, allowing users to input their questions and receive responses audibly. 
The dynamic nature of the tool facilitates an interactive dialogue between the user and the LLM. Users can pose follow-up questions, maintaining the conversation history.
Users can reset the conversation at any time to start a new chat.

Note that before starting the conversation with the AI assistant, the user needs to configure the OpenAI connection by selecting an LLM and entering the corresponding OpenAI API key.

\section{Evaluation}\label{sec:ev}

In this section, we comprehensively evaluate our framework by addressing the first three research questions we defined in \autoref{sec:intro}:
\begin{itemize}
    \item \textbf{Abstraction Methods (RQ1):} In \autoref{sec:ev:1}, we compare the different BPMN abstraction methods supported within our framework.
    \item \textbf{Prompting Strategies (RQ2):} In \autoref{sec:ev:2}, we explore the effect of various prompting strategies on the comprehension of BPMN models by LLMs.
    \item \textbf{State-Of-The-Art LLMs (RQ3):} In \autoref{sec:ev:3}, we assess the overall effectiveness of integrating state-of-the-art LLMs with our framework.
\end{itemize}

\subsection{Setup}
This section describes the experimental setup. First, we detail the dataset comprising diverse BPMN models and user queries in \autoref{sec:ev:setup:data}. Then, we outline our scoring mechanism used to evaluate LLM outputs in \autoref{sec:ev:setup:score}. 

\subsubsection{Dataset}\label{sec:ev:setup:data}

We use three BPMN models in our experiment. First, we use a publicly available BPMN model (\emph{Healthcare Process}) \cite{munoz-gama2019}. This model represents a standard medical process and is publicly available with a textual description. It provides a baseline for comparison but is limited to simple workflow elements.

To ensure an unbiased evaluation and to address the possibility that the healthcare model and its description might have been included in the training data of the LLMs, we designed two additional BPMN models specifically for this study. These models incorporate a broader range of BPMN elements such as events, data objects, and complex gateways, ensuring a comprehensive assessment of the LLMs’ capabilities across various BPMN facets.

The following list gives an overview of the two additional BPMN models we designed for our evaluation together with the help of BPMN experts from our research team: 
\begin{itemize}

    \begin{figure}[!t]
        \centering
        \includegraphics[width=0.9\textwidth]{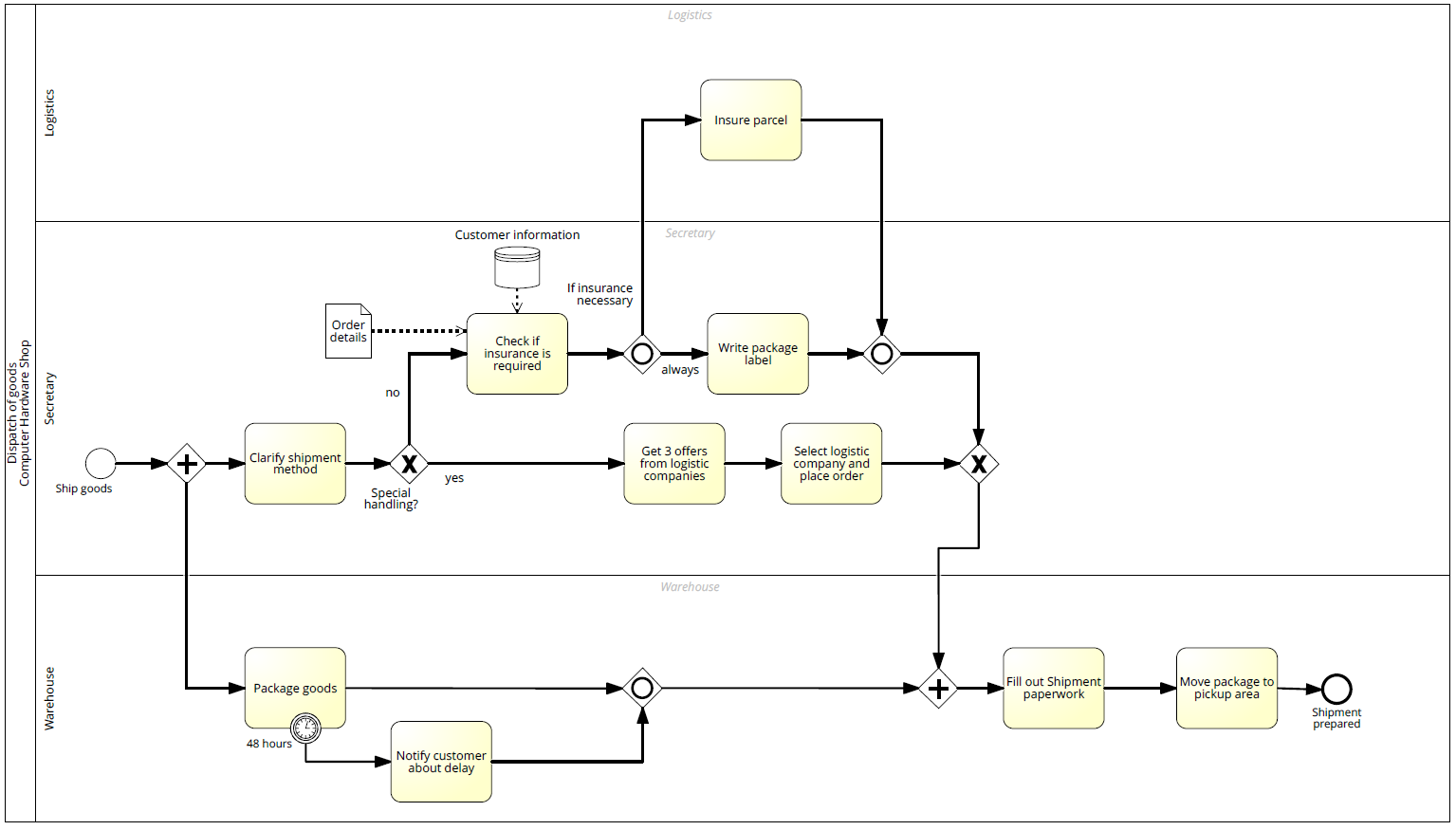}
        \caption{Dispatch Of Goods BPMN model.}
        \label{fig:bpmn:goods}
    \end{figure}

    \item \emph{Dispatch of Goods}: This process, shown in \autoref{fig:bpmn:goods}, details the preparation and dispatch of goods, starting from the decision on shipping methods to the final packaging and shipment preparation. It includes advanced BPMN elements such as time events, data objects, and inclusive gateways.

    \begin{figure}[!t]
        \centering
        \includegraphics[width=\textwidth]{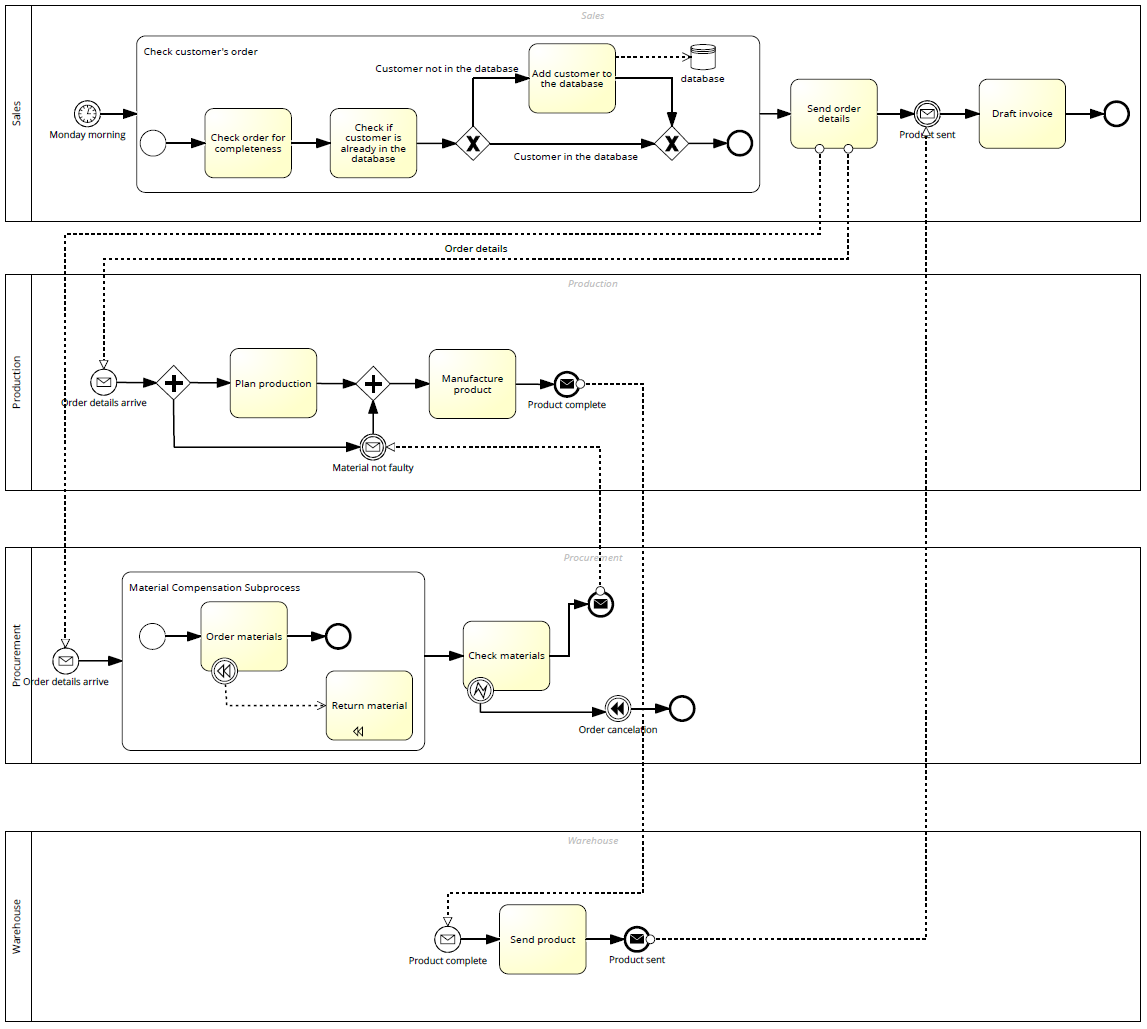}
        \caption{Order Manufacturing BPMN model.}
        \label{fig:bpmn:order}
    \end{figure}

    \item \emph{Order Manufacturing}: This process, shown in \autoref{fig:bpmn:order}, revolves around handling and fulfilling customer orders within a company, engaging multiple departments from sales to warehouse. It captures various advanced BPMN elements such as conditional flows, events (e.g., messages, compensation), and sub-processes.

\end{itemize}

\begin{table}[!t]
\centering
\caption{Types of Questions}
\resizebox{\textwidth}{!}{
\begin{tabular}{|c|c|p{13cm}|}
\hline
\multicolumn{2}{|c|}{\textbf{Type}} & \multicolumn{1}{|c|}{\textbf{Description}} \\ \hline
A1 & Open-Ended & Questions allowing for expansive, subjective responses rather than strict factual answers. \\ \hline
A2 & Close-Ended & Questions prompting definitive answers, such as ``yes'' or ``no'', or leading to specific factual information.  \\ \hline
A3 & Control-Flow & Questions concerning the relationships of tasks and gateways. \\ \hline
A4 & Data-Perspective & Questions related to the relationships between data objects and other artifacts. \\ \hline
A5 & Notation & Questions related to the syntax of the BPMN elements in the context of the given process model. \\ \hline
A6 & Domain-Specific & Questions specific to a domain (e.g., finance, healthcare, manufacturing). \\ \hline
A7 & Organizational-Specific & Questions specific to the practice or policies of an organization. \\ \hline
A8 & Role-Perspective & Questions related to involved actors and their responsibilities. \\ \hline
A9 & Event Relationships & Questions concerning relationships between events and other artifacts. \\ \hline
\end{tabular}
}
\label{tab:questionaspects}
\end{table}

\begin{table*}[!t]
\caption{Inquiries used for the ``Healthcare'' process.}
\label{table:questionsHealthcareProcess}
\centering
\resizebox{\textwidth}{!}{
\begin{tabular}{|m{0.8cm}|m{20cm}|}
\hline
\multicolumn{2}{|c|}{\textbf{Question}} \\
\hline
1 & What are the initial steps taken to prepare for the procedure? \\
\hline
2 & Why is hand washing crucial before proceeding with the operation? \\
\hline
3 & At what stage of the process is the puncture area cleaned? \\
\hline
4 & What is the significance of putting sterile gel before the procedure? \\
\hline
5 & How does ultrasound configuration play into the overall procedure? \\
\hline
6 & What techniques are used to identify the correct anatomical puncture site? \\
\hline
7 & At what point is the decision made regarding the need for anesthetics? \\
\hline
8 & How is the guidewire used in the procedure, and what are its implications? \\
\hline
9 & Is there a verification step to ensure the wire and catheters are correctly positioned? \\
\hline
10 & How do operators decide to widen the pathway for the catheter? \\
\hline
11 & What measures are taken to confirm the flow and reflow in the catheter? \\
\hline
12 & What steps are involved in the removal of the guidewire? \\
\hline
13 & How is the final position of the catheter checked before concluding the procedure? \\
\hline
14 & What are the potential risks if the catheter position is not correctly verified? \\
\hline
15 & Why might there be a need for further intervention after advancing the catheter? \\
\hline
16 & How does the process ensure the sterility and safety of the procedure throughout? \\
\hline
17 & What role does Doppler and compression identification play in the procedure? \\
\hline
18 & How do practitioners decide between using Doppler identification, anatomic identification, and compression identification techniques? \\
\hline
19 & After how many stages is blood return expected, and why is it important? \\
\hline
20 & What are the concluding steps of the procedure, and how is success determined? \\
\hline
\end{tabular}
}
\end{table*}



\begin{table*}[!t]
\caption{Inquiries for the ``Dispatch of Goods'' process.}
\label{table:questionsDispatchOfGoods}
\centering
\resizebox{\textwidth}{!}{
\begin{tabular}{|l|m{10cm}|c|c|c|c|c|c|c|c|c|}
\hline
\multicolumn{2}{|c|}{\textbf{Question}} & \textbf{A1} & \textbf{A2} & \textbf{A3} & \textbf{A4} & \textbf{A5} & \textbf{A6} & \textbf{A7} & \textbf{A8} & \textbf{A9} \\
\hline
 1 & Can the activities ``package goods'' and ``get 3 offers from logistic companies'' be executed in parallel? & ~ & x & x & ~ & ~ & ~ & ~ & ~ & ~ \\ \hline
        2 & Give an explanation for the process steps that the secretary carries out! & x & ~ & x & x & ~ & x & ~ & x & ~ \\ \hline
        3 & Give an explanation for the ``warehouse'' process! & x & ~ & x & ~ & ~ & x & ~ & x & x \\ \hline
        4 & How many activities does Logistics perform? & ~ & x & ~ & ~ & ~ & ~ & ~ & ~ & ~ \\ \hline
        5 & What does the gateway after the activity ``clarify shipment method'' mean? & x & ~ & x & ~ & x & ~ & ~ & ~ & ~ \\ \hline
        6 & When does the logistics department need to insure parcel? & ~ & x & x & ~ & ~ & ~ & x & ~ & ~ \\ \hline
        7 & How does the process end? & ~ & x & x & ~ & ~ & ~ & ~ & ~ & ~ \\ \hline
        8 & How many organizational units are involved in the process? & ~ & x & ~ & ~ & ~ & ~ & ~ & x & ~ \\ \hline
        9 & Can writing the package label be executed simultaneously to selecting a logistic company and placing an order? & ~ & x & x & ~ & ~ & ~ & x & ~ & ~ \\ \hline
        10 & What data does the secretary use when it checks if an insurance is required? & ~ & x & ~ & x & ~ & ~ & x & ~ & ~ \\ \hline
        11 & When does the warehouse have to notify the customer about the delay? & ~ & x & x & ~ & ~ & ~ & x & ~ & x \\ \hline
        12 & Is the activity ``package goods'' finished if 48 hours have passed? & ~ & x & x & ~ & ~ & ~ & ~ & ~ & x \\ \hline
        13 & Can the activities ``insure parcel'' and ``write package label'' be executed simultaneously? & ~ & x & x & ~ & ~ & ~ & ~ & ~ & ~ \\ \hline
        14 & Is it possible that only one of the activities ``insure parcel'' and ``write package label'' is executed? & ~ & x & x & ~ & ~ & ~ & ~ & ~ & ~ \\ \hline
        15 & Which activity includes information retrieval from a database? & ~ & x & ~ & x & ~ & ~ & ~ & ~ & ~ \\ \hline
\end{tabular}
}
\end{table*}

To thoroughly evaluate the quality of answers generated by the LLMs, we generate questions that cover the various aspects shown in \autoref{tab:questionaspects}. These aspects were designed to cover different facets of process understanding, ranging from basic control-flow sequencing to complex data interactions and organizational roles. However, it is important to note that for the healthcare process, we do not have questions that capture all of these aspects as the corresponding model lacks advanced BPMN elements. The questions for the three processes are shown in \autoref{table:questionsHealthcareProcess}, \autoref{table:questionsDispatchOfGoods}, and \autoref{table:questionsOrderManufacturing}.

\begin{table*}[!t]
\caption{Inquiries for the ``Order Manufacturing'' process.}
\label{table:questionsOrderManufacturing}
\centering
\resizebox{\textwidth}{!}{
\begin{tabular}{|l|m{10cm}|c|c|c|c|c|c|c|c|c|}
\hline
\multicolumn{2}{|c|}{\textbf{Question}} & \textbf{A1} & \textbf{A2} & \textbf{A3} & \textbf{A4} & \textbf{A5} & \textbf{A6} & \textbf{A7} & \textbf{A8} & \textbf{A9} \\
\hline
1 & How does the sales process work? & x & ~ & x & x & ~ & x & ~ & x & ~ \\ \hline
        2 & What is the condition for the production to manufacture the product? & ~ & x & ~ & ~ & ~ & ~ & ~ & x & x \\ \hline
        3 & How does the subprocess in the activity ``check customer’s order'' work? & x & ~ & x & x & ~ & ~ & x & ~ & ~ \\ \hline
        4 & Who is responsible for sending the product to the customer? & ~ & x & ~ & ~ & ~ & ~ & x & x & ~ \\ \hline
        5 & What happens after the order details arrive at the procurement team? & x & ~ & x & ~ & ~ & x & ~ & x & x \\ \hline
        6 & What happens if an error occurs when checking the materials? & x & ~ & x & ~ & ~ & ~ & x & ~ & x \\ \hline
        7 & How does a compensation work? & x & ~ & x & ~ & ~ & ~ & ~ & ~ & x \\ \hline
        8 & Why does the activity ``return material'' contain a compensation symbol? & ~ & x & ~ & ~ & x & ~ & ~ & ~ & ~ \\ \hline
        9 & Why does the compensation need its own subprocess? & x & ~ & x & ~ & x & ~ & ~ & ~ & x \\ \hline
        10 & What is the meaning of a message symbol inside of an event? & ~ & x & ~ & ~ & x & ~ & ~ & ~ & x \\ \hline
        11 & What is the difference between a filled event symbol and a non-filled event symbol? & ~ & x & ~ & ~ & x & ~ & ~ & ~ & x \\ \hline
        12 & What does the parallel gateway mean? & ~ & x & x & ~ & x & ~ & ~ & ~ & ~ \\ \hline
        13 & Is the material checked after manufacturing the product? & ~ & x & ~ & ~ & ~ & x & ~ & ~ & x \\ \hline
        14 & Does the plan production occur after manufacture product?  & ~ & x & x & ~ & ~ & x & ~ & ~ & ~ \\ \hline
        15 & Does the sales team check the customer’s order? & ~ & x & ~ & ~ & ~ & x & ~ & x & ~ \\ \hline
        16 & Is the product sent before drafting the invoice? & ~ & x & x & ~ & ~ & ~ & ~ & ~ & ~ \\ \hline
\end{tabular}
}
\end{table*}

For the healthcare process, we utilize the overall textual model description of the underlying process as ground truth. For the other two models, we designed a ground truth answer for each question. 

\subsubsection{Outputs Scoring} \label{sec:ev:setup:score}

In \cite{DBLP:conf/bpmds/BertiKHL024}, evaluation criteria for LLMs' outputs on process mining tasks are proposed. These criteria include \emph{human evaluation}, where a human evaluates the text provided by the LLM; \emph{automatic evaluation}, which is only applicable to quantitative questions; and \emph{self-evaluation}, where LLMs evaluate LLM outputs. Particularly, \emph{self-reflection} methods, where the output of one session is provided to another LLM or another session of the same LLM to score the quality of the output, can be employed for self-evaluation \cite{DBLP:journals/corr/abs-2310-06271}.

For the evaluation of our framework, we decided to employ LLM self-evaluation due to the large size of the experiments. We incorporate a direct comparison of LLM outputs to ground truth answers. We utilize \emph{gpt-4o} to perform a self-evaluation, assigning quality scores to LLM responses based on their alignment with the ground truth answers. To assess the robustness of our evaluation, we conducted a detailed review involving human experts; specifically, we assigned this role to the team who designed the models and the ground truth answers. We selected a sample encompassing two questions from each process, with four different LLM-generated answers for each question. Our experts were also provided with the rationale used by \emph{gpt-4o} for scoring these answers, to assess both the quality and fairness of the evaluation process.

The feedback we received on the LLM self-evaluation has been positive in general, affirming that the quality of the evaluation is high and the reasoning behind the scores is both transparent and well articulated. Although there is an inherent level of subjectivity in any evaluation, our experts agreed that the scores assigned by \emph{gpt-4o} were justified and in line with their understanding of the content. The detailed and contextually rich responses from the LLMs were particularly praised for their relevance. An oversight in detecting an error in one answer was noted; however, it did not detract significantly from the overall success of the evaluation method. In general, our experts confirmed the reliability and high accuracy of the LLM self-evaluation approach. 

\subsection{Results}\label{subsec:res}

The automatic evaluation is reproducible by downloading the code from \url{https://github.com/humam-kourani/AIPA} and executing the scripts contained in the \textbf{offline\_tests/questions\_auto\_eval.py} folder. In particular, \textbf{answer\_questions.py} is used to execute the questions and \textbf{evaluate\_questions.py} can be used to evaluate their results. Each script can be modified with the LLM's connection parameters (API URL, name of the model, and API key) and allows the configuration of the abstraction, prompting strategies, the BPMN model, the list of questions, and the ground truth answers. The BPMN models are stored in the folder \textbf{offline\_tests/bpmn\_models}, and the questions with their ground truth answers are stored in \textbf{offline\_tests/data}. We collect in \textbf{evaluation/evaluation\_results.zip} the evaluation results of all the experiments. For each dataset, we include the different experiments (abstractions, prompting strategies, choice of the LLM) in different folders.

\subsubsection{Evaluating Abstraction Formats}\label{sec:ev:1}

In this section, we perform a comparative evaluation of the different abstraction formats defined in \autoref{sec:abst}. We configure our framework to use \emph{gpt-4o-2024-05-13}, and we only enable simple prompting strategies (S1, S2, S3, and S4) that are efficient in terms of token usage, allowing us to focus on the impact of the abstraction format itself on the comprehension of BPMN models. The results of evaluating the effect of the abstraction formats on LLM comprehension are detailed in \autoref{table:abst}, where we report the average quality scores across various question categories, the overall average quality score, and the number of tokens required for each format.

\begin{table*}[!t]
\centering
\caption{Effect of abstraction formats on the comprehension of BPMN models by \emph{gpt-4o-2024-05-13}. Average quality scores and numbers of required tokens are reported.}
\label{table:abst}

\begin{subtable}{0.95\textwidth}
\caption{Healthcare Process.}
\resizebox{\textwidth}{!}{
\begin{tabular}{|c|cc|cc|cc|cc|}
\hline
 & \multicolumn{2}{c|}{JSON} & \multicolumn{2}{c|}{SXML} & \multicolumn{2}{c|}{XML} & \multicolumn{2}{c|}{PNG} \\
 & Score & \#Tokens & Score & \#Tokens & Score & \#Tokens & Score & \#Tokens \\
\hline
\hspace{0.4cm} All \hspace{0.4cm} & $8.5 \pm 0.8$ & $8918 \pm 2$ & $\mathbf{8.7 \pm 0.6}$ & $7079 \pm 2$ & $8.5 \pm 1.1$ & $26329 \pm 2$ & $4.9 \pm 2.5$ & $1015 \pm 2$\\
\hline
\end{tabular}
}
\end{subtable}

\hfill

\begin{subtable}{0.95\textwidth}
\caption{Dispatch of Goods.}
\resizebox{\textwidth}{!}{
\begin{tabular}{|c|cc|cc|cc|cc|c|}
    \hline
    Question & \multicolumn{2}{c|}{JSON} & \multicolumn{2}{c|}{SXML} & \multicolumn{2}{c|}{XML} & \multicolumn{2}{c|}{PNG}\\
    \cline{2-9}
    Group & Score & \#Tokens & Score & \#Tokens & Score & \#Tokens & Score & \#Tokens \\
    \hline
        A1 & \textbf{8} & 4542 & \textbf{8} & 3829 & 7.83 & 16509 & 7.5 & 1353 \\ 
        A2 & 6.69 & 4544.08 & \textbf{6.73} & 3831.08 & 6.42 & 16511.08 & 6.42 & 1355.08 \\ 
        A3 & \textbf{7.58} & 4544.17 & 7.54 & 3831.17 & 7.13 & 16511.17 & 6.54 & 1355.17 \\ 
        A4 & 7.33 & 4541.67 & 8 & 3828.67 & 7.83 & 16508.67 & \textbf{8.83} & 1352.67 \\ 
        A5 & \textbf{9} & 4545 & 8.5 & 3832 & \textbf{9} & 16512 & 8.5 & 1356 \\
        A6 & 7.5 & 4540.5 & \textbf{7.75} & 3827.5 & 7.25 & 16507.5 & 7 & 1351.5 \\
        A7 & 6.25 & 4544 & 6 & 3831 & 6.25 & 16511 & \textbf{7.38} & 1355 \\ 
        A8 & \textbf{8} & 4540.33 & 8 & 3827.33 & 7.83 & 16507.33 & 7.33 & 1351.33 \\ 
        A9 & 7 & 4543.75 & \textbf{7.38} & 3830.75 & 6.63 & 16510.75 & 6.75 & 1354.75 \\ 
        \hline
        All & $6.9 \pm 2.1$ &	$4544 \pm 5$&	$\mathbf{7.0 \pm 2.0}$&	$3830 \pm 4$&	$6.7 \pm 2.8$&	$16511 \pm 5$	&$6.6 \pm 2.2$	&$1355 \pm 5$\\ \hline

\end{tabular}}
\end{subtable}

\hfill

\begin{subtable}{0.95\textwidth}
\caption{Order Manufacturing.}
\resizebox{\textwidth}{!}{
\begin{tabular}{|c|cc|cc|cc|cc|c|}
    \hline
    Question & \multicolumn{2}{c|}{JSON} & \multicolumn{2}{c|}{SXML} & \multicolumn{2}{c|}{XML} & \multicolumn{2}{c|}{PNG}\\
    \cline{2-9}
    Group & Score & \#Tokens & Score & \#Tokens & Score & \#Tokens & Score & \#Tokens \\
    \hline  
        A1 & 7.5 & 11338.14 & 7.36 & 8170.14 & \textbf{8.21} & 28365.14 & 8.07 & 1011.14 \\
        A2 & 8.45 & 11338.8 & 7.9 & 8170.8 & \textbf{8.85} & 28365.8 & 7.6 & 1011.8 \\ 
        A3 & 7.78 & 11337.33 & 7.72 & 8169.33 & \textbf{8.61} & 28364.33 & 8.22 & 1010.33 \\ 
        A4 & 7 & 11338.5 & 7.75 & 8170.5 & \textbf{8.25} & 28365.5 & \textbf{8.25} & 1011.5 \\
        A5 & 8.4 & 11339.6 & 7.7 & 8171.6 & \textbf{8.6} & 28366.6 & 7.9 & 1012.6 \\ 
        A6 & 8.08 & 11337.67 & 7.58 & 8169.67 & \textbf{8.5} & 28364.67 & 7.42 & 1010.67 \\ 
        A7 & 7.33 & 11340 & 7.5 & 8172 & \textbf{8.5} & 28367 & 8 & 1013 \\
        A8 & 7.67 & 11338.5 & 7.67 & 8170.5 & \textbf{8.67} & 28365.5 & 8.58 & 1011.5 \\ 
        A9 & 8.17 & 11339.11 & 7.22 & 8171.11 & \textbf{8.22} & 28366.11 & 7.06 & 1012.11 \\ 
        \hline
        All & $8.1 \pm 0.9$ &	$11339 \pm 3$ &	$7.7 \pm 1.4$ &	$8171 \pm 3$ &	$\mathbf{8.6 \pm 0.8}$ &	$28365 \pm 3$ &	$7.8 \pm 1.8$ &	$1012 \pm 3$
\\ \hline
\end{tabular}}
\end{subtable}

\end{table*}

Our findings indicate that JSON, simplified XML (SXML), and XML generally produce similar comprehension scores, with PNG lagging behind overall. This difference becomes especially apparent for the healthcare  process, where the PNG format scored very low compared to the other formats. However, PNG showed better performance in answering data-related and role-perspective questions, likely due to its visual nature, which makes artifacts and swimlanes more recognizable. 

XML stands out for the Order Manufacturing process. Its detailed representation leads to superior results across all categories but at the cost of a significantly higher token consumption compared to JSON and SXML.

In summary, the choice of the abstraction format depends on the type of question and the complexity of the processes involved. While XML and PNG each excel in certain scenarios, JSON and SXML consistently offer the best trade-off between quality and token consumption.

\subsubsection{Evaluating Prompting Strategies}\label{sec:ev:2}

In this subsection, we explore the impact of the prompting strategies we defined in \autoref{sec:prompt} on LLMs' comprehension of BPMN models. We employing both the JSON and simplified XML abstraction formats for each BPMN model, and we configure our framework to use \emph{gpt-4o-2024-05-13}. Each prompting strategy is assessed individually to determine its specific effect; however, for practical applications, combining these strategies is recommended to achieve optimal results. The average quality scores and standard deviation values for each strategy are reported in \autoref{tab:ev:promt}.

\begin{table*}[!t]
\caption{Effect of prompting strategies on the comprehension of BPMN models by \emph{gpt-4o-2024-05-13}. For each model, both the JSON and simplified XML abstractions are considered. The average and standard deviation values are reported.}\label{tab:ev:promt}
\centering
\resizebox{\textwidth}{!}{
\begin{tabular}{|c|c|c|c|c|c|c|c|c|c|}
\hline
Model    & Abst. & \textbf{None} & \textbf{S1} & \textbf{S2} & \textbf{S3} & \textbf{S4} & \textbf{S5} & \textbf{S6} & \textbf{S7} \\ \hline
\multirow{2}{*}{Healthcare}    & JSON        & $6.8 \pm 2.3$ & $6.3 \pm 2.5$        & $7.4 \pm 1.9$      & $\mathbf{8.4 \pm 0.9}$ & $6.4 \pm 2.2$   & $7.4 \pm 2.0$     & $7.3 \pm 2.1$          & $\mathbf{8.3 \pm 1.1}$  \\
 & SXML        & $7.4 \pm 2.0$ & $7.5 \pm 1.9$        & $7.5 \pm 2.1$      & $\mathbf{8.7 \pm 0.7}$ & $7.1 \pm 2.1$   & $6.8 \pm 2.5$     & $7.0 \pm 2.5$          & $\mathbf{8.4 \pm 1.1}$  \\ \hline

\multirow{2}{*}{Order}    & JSON        & 6.0 $\pm$ 2.1 & 6.2 $\pm$ 2.1        & 6.4 $\pm$ 1.5      & $\mathbf{7.7 \pm 1.4}$ & 6.4 $\pm$ 1.8   & 6.2 $\pm$ 1.8     & $\mathbf{6.9 \pm 1.4}$ & $\mathbf{7.2 \pm 1.6}$  \\
    & SXML        & 5.8 $\pm$ 1.6 & 6.0 $\pm$ 1.6        & 5.6 $\pm$ 1.7      & $\mathbf{8.1 \pm 1.0}$ & 6.3 $\pm$ 1.4   & 6.1 $\pm$ 1.9     & $\mathbf{6.9 \pm 1.6}$ & $\mathbf{7.1 \pm 1.5}$  \\ \hline
    
\multirow{2}{*}{Dispatch} & JSON        & $6.0 \pm 1.9$ & $5.5 \pm 1.9$        & $5.9 \pm 1.6$      & $\mathbf{6.6 \pm 2.0}$ & $5.6 \pm 1.7$   & $5.7 \pm 1.8$     & $\mathbf{6.8 \pm 2.0}$ & $6.5 \pm 2.1$           \\
 & SXML        & $5.7 \pm 1.3$ & $6.1 \pm 2.1$        & $6.2 \pm 1.6$      & $\mathbf{7.2 \pm 1.8}$ & $5.6 \pm 1.5$   & $5.1 \pm 1.6$     & $\mathbf{6.7 \pm 1.9}$ & $5.7 \pm 2.5$ \\ \hline
\end{tabular}}
\end{table*}


The results show that strategies S3 (non-technical restriction), S6 (few-shot learning), and S7 (negative prompting) stand out with particularly strong positive impacts across the different models and abstraction formats. For S3, restricting LLMs to use only natural language significantly enhances model comprehension, likely due to improved readability and accessibility of the information. Strategies S6 and S7 also align with expectations, showing a positive effect and confirming the usefulness of both positive and negative example-based learning for enhancing LLM comprehension.

Conversely, strategies S1 and S2, which involve role prompting, demonstrate minimal impact on the comprehension capabilities of LLMs. This outcome suggests that simple identity-based prompts do not significantly influence LLM performance in this context. Similarly, the chain-of-thoughts approach (S4) and the knowledge injection strategy (S5) show limited effects. For S4, the tendency to generate unnecessarily lengthy outputs might overwhelm the user with information that does not directly aid in model comprehension. For S5, the lack of observed improvement might be attributed to the extensive availability of information on the BPMN 2.0 standard online, which the LLM could have already encountered during its training. Future work could explore more targeted knowledge injection strategies, focusing specifically on the complex elements of the given BPMN model rather than providing general knowledge.

\subsubsection{Evaluating State-Of-The-Art LLMs}\label{sec:ev:3}




In this section, we evaluate the performance of state-of-the-art LLMs on our framework. We configure our framework to use the Simplified XML (SXML) abstraction, and we enable the first four prompting strategies (S1, S2, S3, and S4). We have selected four LLMs for this analysis: 

\begin{itemize}
    \item \textbf{GPT-4o-2024-05-13}: The latest iteration of OpenAI's language models\footnote{\url{https://openai.com/}}, known for its broad and extensive training on a diverse dataset which may enhance its capability in domain-specific questions. The context window of this LLM is {\bf 128K}. It also supports visual prompts.
    \item \textbf{GPT-4-Turbo-2024-04-09}: A variant of GPT-4 optimized for speed and efficiency, potentially sacrificing some depth in exchange for faster response times. The context window of this LLM is {\bf 128K}. It also supports visual prompts. On a note, this LLM is outperformed in most known benchmarks by GPT-4o in both response time and quality.
    \item \textbf{Microsoft WizardLM-2-8x22B} \footnote{\url{https://ollama.com/library/wizardlm2:8x22b}}: An open-source high-capacity model from Microsoft, designed to excel in deep contextual understanding and complex reasoning tasks. The context window of this LLM is {\bf 64K}, still allowing for the full provision of our considered textual prompts.
    \item \textbf{Mixtral-8x22B-Instruct v0.1} \footnote{\url{https://ollama.com/library/mixtral:8x22b-instruct-v0.1-fp16}}: An open-source instructive model geared towards following explicit user instructions with high precision, using fewer tokens for outputs which might affect the depth of generated content. The context window of this LLM is {\bf 64K}, still allowing for the full provision of our considered textual prompts.
\end{itemize}

We also considered smaller LLMs (\textbf{Mixtral-8x7B-Instruct v0.1}\footnote{\url{https://ollama.com/library/mixtral:8x7b-instruct-v0.1-fp16}}, \textbf{Codestral}\footnote{\url{https://ollama.com/library/codestral:22b-v0.1-f16}}, \textbf{Mistral 7B}\footnote{\url{https://ollama.com/library/mistral:7b-instruct-fp16}}) in our initial experiments, but in light of significantly worse results, we discarded them. Some other LLMs were discarded for their limited context window (\textbf{Llama3-70B}\footnote{\url{https://ollama.com/library/llama3:70b-instruct-fp16}}, \textbf{Nemotron-4-340B-Instruct}\footnote{\url{https://deepinfra.com/nvidia/Nemotron-4-340B-Instruct}}).

The results of our comparative analysis are shown in \autoref{table:llms}, where we report the average quality scores for different question categories along with the number of output tokens produced by each LLM. 

\begin{table*}[!t]
\centering
\caption{Performance of state-of-the-art LLMs on the comprehension of BPMN models. Average quality scores and average numbers of output tokens are reported.}
\label{table:llms}

\begin{subtable}{0.95\textwidth}
\caption{Healthcare Process.}
\resizebox{\textwidth}{!}{
\begin{tabular}{|c|cc|cc|cc|cc|}
\hline
\multirow{2}{*}{} & \multicolumn{2}{c|}{gpt-4o} & \multicolumn{2}{c|}{gpt-4-turbo} & \multicolumn{2}{c|}{WizardLM-2-8x22B} & \multicolumn{2}{c|}{Mixtral-8x22B-Instruct} \\
 & Score & Output Tok. & Score & Output Tok. & Score & Output Tok. & Score & Output Tok. \\
\hline
\hspace{0.4cm} All \hspace{0.4cm} & $\mathbf{8.7 \pm 0.6}$ & $408 \pm 168$ & $8.6 \pm 0.8$ & $319 \pm 108$ & $8.5 \pm 1.1$ & $430 \pm 110$ & $8.3 \pm 1.0$ & $154 \pm 94$ \\
\hline
\end{tabular}
}
\end{subtable}

\hfill

\begin{subtable}{0.95\textwidth}
\caption{Dispatch of Goods.}
\resizebox{\textwidth}{!}{
\begin{tabular}{|c|cc|cc|cc|cc|c|}
\hline
Question & \multicolumn{2}{c|}{gpt-4o} & \multicolumn{2}{c|}{gpt-4-turbo} & \multicolumn{2}{c|}{WizardLM-2-8x22B} & \multicolumn{2}{c|}{Mixtral-8x22B-Instruct} \\
\cline{2-9}
Group & Score & Output Tok. & Score & Output Tok. & Score & Output Tok. & Score & Output Tok. \\
\hline
        A1 & \textbf{8} & 504.33 & 7.5 & 414.33 & 7.67 & 448 & 7.33 & 222 \\ 
        A2 & 6.73 & 263.54 & 6.31 & 267.38 & \textbf{7.31} & 363.62 & 6.5 & 146.08 \\ 
        A3 & 7 & 352.25 & 6.54 & 347.08 & \textbf{7.46} & 415.67 & 6.58 & 173 \\ 
        A4 & \textbf{8} & 393 & 7.67 & 279.33 & 7.33 & 390.67 & 6.33 & 184.67 \\ 
        A5 & \textbf{8.5} & 330 & 8 & 200 & 8 & 320 & 8 & 135 \\
        A6 & \textbf{7.75} & 591.5 & 7.25 & 521.5 & 7.5 & 512 & 7 & 265.5 \\ 
        A7 & 6 & 246 & 5.5 & 219 & \textbf{6.75} & 411.75 & 6.5 & 148.25 \\ 
        A8 & \textbf{8} & 434.33 & 7 & 407.67 & \textbf{8} & 440.67 & 7.83 & 221.67 \\ 
        A9 & \textbf{7.38} & 366 & 7.13 & 374 & 7.25 & 428 & 6.63 & 199.75 \\ 
        \hline
        All & $7.0 \pm 2.0$ & $309 \pm 160$ & $6.5 \pm 1.9$ & $295 \pm 172$ & $\mathbf{7.4 \pm 1.2}$ & $371 \pm 110$ & $6.7 \pm 1.7$ & $160 \pm 86$ \\ 
        \hline
\end{tabular}
}
\end{subtable}

\hfill

\begin{subtable}{0.95\textwidth}
\caption{Order Manufacturing.}
\resizebox{\textwidth}{!}{
\begin{tabular}{|c|cc|cc|cc|cc|c|}
\hline
Question & \multicolumn{2}{c|}{gpt-4o} & \multicolumn{2}{c|}{gpt-4-turbo} & \multicolumn{2}{c|}{WizardLM-2-8x22B} & \multicolumn{2}{c|}{Mixtral-8x22B-Instruct} \\
\cline{2-9}
Group & Score & Output Tok. & Score & Output Tok. & Score & Output Tok. & Score & Output Tok. \\
\hline
        A1 & 7.36 & 449.86 & \textbf{7.64} & 406.57 & 7 & 501.43 & 7.07 & 163.29 \\
        A2  & \textbf{7.9} & 330.1 & \textbf{7.9} & 266.3 & 7.75 & 345.1 & 7.1 & 130.4 \\
        A3 & \textbf{7.72} & 403.22 & 7.56 & 364.33 & 7.67 & 466.33 & 7.5 & 152.78 \\
        A4 & 7.75 & 512.5 & 7 & 411.5 & \textbf{8.5} & 475 & 8 & 158.5 \\
        A5 & 7.7 & 433.2 & \textbf{8.2} & 353.2 & 7.2 & 411.2 & 6.6 & 179.6\\
        A6 & \textbf{7.58} & 354.33 & \textbf{7.58} & 308 & 7.25 & 412.17 & 7.08 & 116.5 \\
        A7 & 7.5 & 300.67 & 7.17 & 321 & 7.33 & 408 & \textbf{7.67} & 167.33 \\
        A8 & 7.67 & 347.67 & \textbf{7.75} & 322 & 7.33 & 358.5 & 7.08 & 126.83 \\
        A9 & 7.22 & 408.22 & \textbf{7.89} & 370.78 & 6.78 & 458.78 & 6.39 & 158.78 \\
        \hline
        All & $7.7 \pm 1.4$ & $379 \pm 132$ & $\mathbf{7.8 \pm 0.9}$ & $324 \pm 119$ & $7.4 \pm 1.5$ & $409 \pm 115$ & $7.1 \pm 1.5$ & $144 \pm 53$\\ 
        \hline
\end{tabular}
}
\end{subtable}
\end{table*}

All LLMs demonstrated similar levels of performance; however, \textbf{GPT-4o-2024-05-13} stands out in domain-specific questions, possibly due to its more extensive training on a wider range of topics and scenarios. This suggests that a broader training corpus can significantly benefit domain-specific comprehension.

Notably, the results show that open-source LLMs perform comparably to commercial alternatives, highlighting the advancements and accessibility in LLM technology.

Interestingly, there is a consistent trend where increased token usage correlates with better scores. This suggests that more detailed responses, which utilize more tokens, tend to provide richer and more useful information, thereby improving the model's score in our evaluation. Among the evaluated models, \textbf{Mixtral-8x22B-Instruct} uses the fewest tokens but maintains competitive performance, indicating efficiency in token usage. However, the detailed, token-rich outputs from models like \textbf{GPT-4o} and \textbf{WizardLM} indicate that when the depth of description is critical, especially for complex topics, higher token consumption may be beneficial.

\section{User Study}\label{sec:study}
In this section, we present a user study designed to assess the usability and effectiveness of our LLM-based framework for process model comprehension in real-world contexts. The study specifically addresses RQ4 (cf. \autoref{sec:intro}), exploring how industry experts in the field of business process management perceive and interact with BPMN models using our tool, AIPA. It is important to note that the version of AIPA used in this study did not include the voice support functionalities, as these features were added subsequently. The user study was conducted by configuring AIPA to use \emph{gpt-4-1106-preview}.

\subsection{Setup}
We conducted a total of six interviews, each lasting between 45 minutes and one hour. The interview partners were contacted via email and selected for their extensive experience with BPMN modeling in professional contexts and their background knowledge of how LLMs work. The experts were identified through the professional networks of the authors, who are well-connected within the BPM research and industry community. The selected experts represent a diverse range of professional experiences, from 0-5 years to 15-20 years in the field. All invited experts agreed to participate in the user study. \autoref{tab:partnersInvolvedCaseStudy} provides an overview of the interview partners.

\begin{table*}[!t]
\caption{Partners involved in the user-study.}
\label{tab:partnersInvolvedCaseStudy}
\resizebox{\textwidth}{!}{
\begin{tabular}{|l|l|l|l|}
\hline
\textbf{Partner} & \textbf{Role}                  & \textbf{Company} & \textbf{Experience}                                   \\ \hline
\textbf{Expert 1}                & Practice Lead Green BPM        & Consultancy for Sustainable IT Applications & 15-20 years        \\ \hline
\textbf{Expert 2}                & Green BPM Consultant           & Consultancy for Sustainable IT Applications & 0-5 years        \\ \hline
\textbf{Expert 3}                & Product and Innovation Manager & Software Developer for Process Automation Solutions & 5-10 years\\ \hline
\textbf{Expert 4}                & Co-Founder                     & Process Automation and Software Development & 5-10 years        \\ \hline
\textbf{Expert 5}                & CoE Lead Process Mining        & Automotive Manufacturer    & 0-5 years                         \\ \hline
\textbf{Expert 6}                & BPM Expert                     & Semiconductor Manufacturer      & 5-10 years                    \\ \hline
\end{tabular}
}
\end{table*}

The user study comprised six semi-structured interviews guided by a 16-question interview guide. Conducting semi-structured expert interviews allowed us to focus on the research topic while gathering in-depth information \cite{DBLP:journals/iando/SchultzeA11}. 

The interview guide is divided into three sections. The first section consists of three questions aimed at understanding the interviewee's expertise and creating a comfortable interview atmosphere, as these questions are easy to answer. 



Next, we presented AIPA, explained its functionality, and gave the interviewees enough time to familiarize themselves with the interface and the selected BPMN model. We chose a BPMN model of low complexity to facilitate quick understanding and interaction. Once they indicated that they had understood the model, we asked the interviewees to ask the AI assistant questions about the BPMN model displayed. On average, four questions were asked by each interviewee. 

After the user had time to interact with the tool, we moved on to the second section of the interview guide. The first question focused on evaluating the user interface, while the remaining eleven questions were aimed at evaluating the responses. These eleven questions are structured using eleven criteria derived from the study \cite{DBLP:conf/fat/SokolF20} which utilized these criteria to evaluate the usability of explainable AI systems. For each question, the interviewees were first asked to provide a rating on a five-point scale about the degree of fulfillment of the criterion and then to explore the reasons for the rating. By incorporating these tested criteria, we have followed a robust evaluation framework that improves the functionality of our system and ensures its usability. 

The following list shows the eleven criteria from \cite{DBLP:conf/fat/SokolF20} we used to define the questions of the second phase of the interview guide: 

\begin{enumerate}[label=\textbf{(C\arabic*)}, leftmargin=2cm]

    \item \textbf{Soundness:} Ensures that the explanations accurately reflect the operations of the underlying model, emphasizing the truthfulness of the information presented
    \item \textbf{Completeness:} Measures whether the explanation covers enough context and cases to be useful across various scenarios, not just the specific instance it was generated for.
    \item \textbf{Contextfullness:} Provides explanations with sufficient background to allow users to understand them in relation to their broader operational environment.
    \item \textbf{Interactiveness:} Allows users to engage with the system dynamically, adjusting and querying the explanations to better suit their understanding or needs.
    \item \textbf{Actionability:} Ensures that the explanations guide users towards meaningful actions based on the insights provided.
    \item \textbf{Chronology:} Considers the timing of events or data points in explanations, highlighting the most recent or relevant information that impacts the outcome.
    \item \textbf{Coherence:} Focuses on the consistency of explanations with the user’s prior knowledge and expectations, making them intuitively easier to accept.
    \item \textbf{Novelty:} Ensures that the information provided is insightful, revealing new information that can provoke thought or lead to unexpected insights.
    \item \textbf{Complexity:} Addresses the level of detail in the explanations, ensuring they are neither too simple to be trivial nor too complex to understand.
    \item \textbf{Personalization:} Tailors explanations to the background knowledge and specific needs of individual users, enhancing their relevance and accessibility.
    \item \textbf{Parsimony:} Emphasizes the brevity of explanations, ensuring they are concise and avoid overloading the user with unnecessary information.

\end{enumerate}


After answering the twelve questions from the second section, we concluded the interview with an open question that gave the interviewee the opportunity to address aspects not covered by the previous questions.



\subsection{Results} 
This section provides a comprehensive summary of the interview results, offering insights and key findings derived from the discussions.

\paragraph*{User-Friendliness}
The user-friendliness aspect of the tool is highly praised as it is clear and concise, making it easy for users to interact with the system (Experts 1, 4). Users appreciate that the BPMN model is visible and can be reset, but they suggest further improvements such as displaying multiple process models at the same time to be able to simultaneously analyze several business processes that are connected with each other and improving visual elements, such as adapting the size of the window that contains the model (Expert 3). The ability to drag elements of the BPMN model directly on the user interface is highlighted as positive (Expert 5). 
It is suggested to further integrate the possibility of enlarging the model view to allow better interaction (Experts 5, 6) and possibly implement a function that highlights in color which part of the model the AI assistant's response refers to (Expert 6). 

\paragraph*{Soundness (C1)}
Regarding the question of soundness, the explanations provided by the AI assistant were generally effective and answered most questions well, although sometimes the answers could have been more precise (Experts 1, 3, 4). The AI assistant explained elements such as message flow accurately, although its impact on the workflow engine was somewhat unclear (Expert 3). It performed well in explaining the exclusive gateway and notations (Expert 3). Overall, more than $80\%$ of the information was correct (Expert 5). However, some issues were also uncovered, such as the AI assistant confusing signals and messages, indicating the need for careful use of correct terminology where a message should be consistently referred to as a message and not a signal (Expert 4). In addition, the AI assistant initially omitted an activity when describing the minimum process and only correctly identified the ``draft invoice'' activity after repeated requests.

\paragraph*{Completeness (C2)}
In response to the question regarding the completeness of the answers, the feedback shows that the explanations are well-generalized and applicable to different elements of BPMN (Experts 1 - 6). The answers are effective for common processes such as the widely used ordering process and can be appropriately applied to similar scenarios. However, for unique or less typical processes, more scrutiny seems required to ensure the accuracy and applicability of the explanations (Expert 1). The AI assistant's ability to provide relevant, overarching explanations suggests that it has an appropriate level of generalizability for practical use in BPMN modeling.

\paragraph*{Contextfullness (C3)}
For the question on contextual fulfillment, the AI assistant received mixed reviews with scores ranging from 3 to 5. While the AI assistant provided detailed and comprehensive answers that improved users' understanding of BPMN processes, concerns were raised that the AI assistant could mislead those unfamiliar with BPMN by relying too much on the AI assistant's explanations (Experts 1, 3). Users appreciated the insights relevant to process optimization, although Expert 5 perceived some responses as confusing, indicating the need for clearer and more precise contextual understanding. There would be further potential for improvement if the AI assistant could access a broader database for more specific analysis (Expert 5).

\paragraph*{Interactiveness (C4)}
Regarding the question on interactivity, the performance of the tool was rated highly and predominantly scored between 4 and 5. It responded effectively to critical suggestions and responded appropriately to follow-up questions. Users appreciated that the AI assistant could recall previous interactions (Expert 3). Although the AI assistant sometimes needed several attempts to give the correct answer (Expert 5), the responsiveness of the AI assistant to user queries was rated positively, which increases interactivity (Experts 2, 3, 6).

\paragraph*{Actionability (C5)}
In addressing actionability, the AI assistant's performance varied, with ratings from 3 to 5. It was praised for providing valuable initial optimization insights that encouraged critical thinking and innovation (Expert 1, 2). Although some suggestions were too general (Expert 3), the AI assistant provided useful technical advice. Technical suggestions were beneficial; however, improvements in modeling and syntax could be better (Expert 6).

\paragraph*{Chronology (C6)}
When asked about chronology, the AI assistant's ability to explain the chronological sequence of events in BPMN models was rated highly and generally received a rating of 5. It explained the sequence of events in different scenarios very effectively (Experts 1 – 6). Overall, the chronology was correctly followed in $100\%$ of cases, although it should be noted that the scores were based on relatively simple models. It was pointed out that the AI assistant could reach its limits with more complex models (Experts 3, 5, 6). 

\paragraph*{Coherence (C7)}
In the assessment of coherence, the AI assistant scored well, with recommendations for clearer definitions of BPMN elements such as gateways and better clarity of process dependencies (Expert 1).
Minor inaccuracies in the explanation of message flows and naming conventions were noted but did not significantly affect the overall coherence (Expert 3).
These limitations emphasize the dependence of the LLM on its training data (Expert 4). Overall, the responses were logical and consistent and matched well with the available information.

\paragraph*{Novelty (C8)}
In terms of novelty, the responses were generally not surprising (Experts 1, 5), but contained practical insights, e.g., on outsourcing activities (Expert 1). Although feedback on portfolio management was appreciated, the responses did not particularly impress those with extensive prior knowledge, suggesting that the novelty of the AI assistant's information may depend on the user's familiarity with the topic (Expert 5).

\paragraph*{Complexity (C9)}
Evaluating complexity, the AI assistant mostly received high marks for the clarity and simplicity of its explanations (Experts 1, 3, 5). The answers were rated as very understandable and were generally in simple language.
However, one criticism was that some answers were initially too long (Experts 4, 6), although the AI assistant was able to provide shorter answers when requested.
It was also suggested that the lists within the responses should be formatted into paragraphs to improve readability and prevent the text from being overwhelming (Expert 6). Moreover, Expert 5 suggested that the answers should be revealed gradually, allowing users to begin reading while the response is still being generated.

\paragraph*{Personalization (C10)}
In the evaluation of personalization, the AI assistant received mixed reviews. It received a high score for accurately addressing the most important use cases but was criticized for needing more diverse examples to improve its application (Expert 4).
While the AI assistant showed good detail orientation, one expert noted that his knowledge exceeded the AI assistant's capabilities, indicating room for improvement in adapting to the experts' expectations (Expert 5).


\paragraph*{Parsimony (C11)}
In the assessment of parsimony, the AI assistant was praised for its precision, particularly in the definition of technical terms and processes such as swim lanes, gateways, and data exchange, which are represented by dashed lines (Experts 1, 3).
However, the answers were often felt to be too long. Suggestions for improvement included starting with shorter initial answers and only providing more detailed follow-up answers on request (Expert 4). While the explanations were generally precise, especially in terms of syntax, it was found that more open-ended questions led to less precise, detailed answers (Expert 6). Overall, while the AI assistant performed well in terms of comprehensiveness, there was a consensus that brevity could be improved.

\paragraph*{Further Feedback}
Answers to the open-ended question in the third section were directed at the need for the AI assistant to provide its responses more quickly and to structure the responses in paragraphs for ease of reading (Expert 1).
It was suggested that the AI assistant should clarify when its answers are merely suggestions that require further review and highlight the specific parts of BPMN that it is addressing. 
Users also wanted more interactive features, such as the ability to adapt the BPMN model directly through the tool to improve the user experience through more dynamic and human-like interactions (Expert 2).
There was a desire for the tool to be able to enable the direct implementation of suggestions for improvement and interactive adjustments to the model (Expert 3).

\section{Discussion and Future Directions}\label{sec:future}

Our approach has demonstrated promising results in using LLMs to enhance the comprehension of process models, yet it also has limitations. This section outlines potential ideas for further research and development.

\emph{Aim for more concise outputs:}
The experts involved in the user study highlighted a tendency of \emph{gpt-4} to produce long textual outputs. A more focused output is preferable to speed up the analytical reasoning. This could be addressed by changing the underlying LLM (for example, \emph{Mixtral-8x22B-Instruct} produces significantly more concise outputs than other considered LLMs) or adopting new prompting strategies explicitly requesting the LLM to produce concise outputs. Also, it is possible to consider training/fine-tuning an LLM to align it with the desired output length. 

\emph{Targeted prompt engineering:}
Despite the promising results of our experimentation, users noticed that LLMs sometimes fail to focus on the correct elements of the BPMN model given the inquiry. The size and complexity of the BPMN abstraction and the additional injected knowledge are overwhelming for the prompt attention mechanisms of current state-of-the-art LLMs. A possible solution is implementing targeted prompt engineering techniques, which involve focusing solely on the specific elements present in the uploaded process models or the user's query. For instance, a Retrieval-Augmented Generation (RAG) pipeline \cite{DBLP:conf/nips/LewisPPPKGKLYR020} could be implemented to store knowledge about different BPMN elements and retrieve the ones relevant for the uploaded model. Similarly, different types of few-shot examples can be stored, and the user's query can be analyzed to only retrieve relevant ones. However, the effectiveness of the retrieval is crucial and it should be thoroughly evaluated when integrated into the framework.

\emph{Providing process knowledge:} LLMs are trained on a vast corpus of data, which allows them to be a general-purpose conversational interface. They also show good domain knowledge in popular benchmarks \cite{DBLP:conf/aaai/GuZYZWZJXLWHXHL24}. However, LLMs may not know in detail how a given process has been implemented in a given setting. Therefore, a user inquiry might be enriched with relevant contextual information about the process. This can be implemented using a RAG pipeline, or by fine-tuning the LLM on specific process knowledge.

\emph{Enhanced interactivity:}
Our framework utilizes LLMs to produce a textual answer to the inquiry of the user. However, analysts would benefit from a feedback mechanism in which the parts of the BPMN model mentioned in the answer are visually highlighted. This would require the specification of a clear output structure to the LLM, in which the textual answer and the configuration of the highlighting are the components. Furthermore, experts from the user study suggested empowering users to implement suggestions for improvement and make interactive adjustments to the BPMN model directly within AIPA. Such features could foster a more responsive and engaging user experience.

\emph{What if one does not know what to ask?} LLMs prove useful in answering user inquiries over a BPMN model. However, also thinking about which inquiries should be asked in order to achieve a complete understanding of the process requires significant expertise by the user. This could be addressed within our framework by integrating a mechanism for automated hypothesis generation, which can also be based on LLMs. This mechanism could help in identifying a list of relevant questions given the process model and the conversational history.

\emph{What if the process is not modeled in BPMN?} The proposed framework is useful for comprehending BPMN models. However, many organizational processes are expressed in textual documents. To use our framework on such processes, \emph{process modeling} could be applied to get a BPMN model from the textual representation \cite{DBLP:conf/bpmds/KouraniB0A24,DBLP:journals/corr/abs-2403-04327}. Integrating process modeling and comprehension techniques would allow process analysts to fully understand the entire set of processes of an organization, even the ones not modeled in BPMN. 

\emph{Integration with process mining:}
In future work, we aim to integrate our LLM-based process model comprehension framework with process mining methodologies \cite{DBLP:books/sp/Aalst16}. In particular, we plan to employ LLMs to interpret and analyze process models that are automatically discovered from event data and annotated with performance metrics or compliance deviations. This approach will bridge the gap between theoretical models and actual process executions, aligning our developments with real-world process dynamics.

\emph{Exploring further LLM applications in BPM:} As we continue to expand the capabilities of LLMs within the BPM field, their potential extends beyond improving process model comprehension. For example, there is a potential to apply LLMs for process enhancement. Future research could investigate how LLMs can not only analyze but also recommend optimizations or variations in process flows.

\section{Conclusion}\label{sec:conc}
In this paper, we introduced a novel framework that utilizes the advanced natural language processing capabilities of Large Language Models (LLMs) to enhance the comprehension of complex process models. We transform intricate Business Process Model and Notation (BPMN) diagrams into various abstraction formats suitable for LLM interpretation, and we explore various prompting strategies to optimize LLM performance. Furthermore, we presented AIPA (AI-Powered Process Analyst), a tool developed to integrate our framework with OpenAI' LLMs. AIPA supports dynamic interactions with process models, enabling users to query BPMN models and receive explanations seamlessly.

We extensively evaluated our framework with different BPMN abstractions and prompting strategies. Our evaluation results indicate that the right combination of model abstraction and prompting strategies significantly improves the model's comprehensibility without compromising detail or accuracy. We conducted a user study to assess the ease of comprehension, accuracy of interpretation, and user satisfaction when interacting with BPMN models through AIPA. Results demonstrate a marked improvement in understanding complex models with our LLM-based framework, confirming its effectiveness and user-friendliness. This paper not only underscores the potential of LLMs to make BPMN models more accessible but also provides empirical evidence, highlighting the transformative potential of leveraging AI technologies for BPM and process mining tasks. 

%


 \bibliographystyle{elsarticle-num} 
 \bibliography{cas-refs}

\begin{thebibliography}{10}
\expandafter\ifx\csname url\endcsname\relax
  \def\url#1{\texttt{#1}}\fi
\expandafter\ifx\csname urlprefix\endcsname\relax\def\urlprefix{URL }\fi
\expandafter\ifx\csname href\endcsname\relax
  \def\href#1#2{#2} \def\path#1{#1}\fi

\bibitem{DBLP:books/sp/DumasRMR18}
M.~Dumas, M.~L. Rosa, J.~Mendling, H.~A. Reijers, Fundamentals of Business
  Process Management, Second Edition, Springer, 2018.

\bibitem{DBLP:books/el/15/RosingWCM15}
M.~von Rosing, S.~White, F.~Cummins, H.~de~Man, Business process model and
  notation - {BPMN}, in: M.~von Rosing, H.~von Scheel, A.~Scheer (Eds.), The
  Complete Business Process Handbook: Body of Knowledge from Process Modeling
  to BPM, Volume {I}, Morgan Kaufmann/Elsevier, 2015, pp. 429--453.
\newblock \href {https://doi.org/10.1016/B978-0-12-799959-3.00021-5}
  {\path{doi:10.1016/B978-0-12-799959-3.00021-5}}.

\bibitem{DBLP:conf/caise/MuehlenR08}
M.~zur Muehlen, J.~Recker, How much language is enough? theoretical and
  practical use of the business process modeling notation, in: Z.~Bellahsene,
  M.~L{\'{e}}onard (Eds.), Advanced Information Systems Engineering, 20th
  International Conference, CAiSE 2008, Montpellier, France, June 16-20, 2008,
  Proceedings, Vol. 5074 of Lecture Notes in Computer Science, Springer, 2008,
  pp. 465--479.
\newblock \href {https://doi.org/10.1007/978-3-540-69534-9\_35}
  {\path{doi:10.1007/978-3-540-69534-9\_35}}.

\bibitem{DBLP:books/daglib/p/MuehlenR13}
M.~zur Muehlen, J.~Recker, How much language is enough? theoretical and
  practical use of the business process modeling notation, in: J.~A.~B. Jr.,
  J.~Krogstie, O.~Pastor, B.~Pernici, C.~Rolland, A.~S{\o}lvberg (Eds.),
  Seminal Contributions to Information Systems Engineering, 25 Years of CAiSE,
  Springer, 2013, pp. 429--443.
\newblock \href {https://doi.org/10.1007/978-3-642-36926-1\_35}
  {\path{doi:10.1007/978-3-642-36926-1\_35}}.

\bibitem{DBLP:journals/bpmj/Recker10}
J.~Recker, Opportunities and constraints: the current struggle with {BPMN},
  Bus. Process. Manag. J. 16~(1) (2010) 181--201.

\bibitem{DBLP:journals/jcis/Bera12}
P.~Bera, Does cognitive overload matter in understanding {BPMN} models?, J.
  Comput. Inf. Syst. 52~(4) (2012) 59--69.

\bibitem{DBLP:journals/corr/abs-2303-08774}
OpenAI, {GPT-4} technical report, CoRR abs/2303.08774 (2023).
\newblock \href {http://arxiv.org/abs/2303.08774} {\path{arXiv:2303.08774}},
  \href {https://doi.org/10.48550/ARXIV.2303.08774}
  {\path{doi:10.48550/ARXIV.2303.08774}}.

\bibitem{DBLP:journals/corr/abs-2303-18223}
W.~X. Zhao, K.~Zhou, J.~Li, T.~T. et~al., A survey of large language models,
  CoRR abs/2303.18223 (2023).

\bibitem{DBLP:conf/aaai/GuZYZWZJXLWHXHL24}
Z.~Gu, X.~Zhu, H.~Ye, L.~Z. et~al., Xiezhi: An ever-updating benchmark for
  holistic domain knowledge evaluation, in: M.~J. Wooldridge, J.~G. Dy,
  S.~Natarajan (Eds.), {IAAI} 2024, {AAAI} Press, 2024, pp. 18099--18107.

\bibitem{DBLP:conf/acl/0009C23}
J.~Huang, K.~C. Chang, Towards reasoning in large language models: {A} survey,
  in: A.~Rogers, J.~L. Boyd{-}Graber, N.~Okazaki (Eds.), Findings of the
  Association for Computational Linguistics: {ACL} 2023, Toronto, Canada, July
  9-14, 2023, Association for Computational Linguistics, 2023, pp. 1049--1065.

\bibitem{zhou_business_2023}
C.~Zhou, D.~Zhang, D.~Chen, C.~Liu, Business {Process} {Complexity}
  {Measurement}: {A} {Systematic} {Literature} {Review}, IEEE Access 11 (2023)
  47940--47955.
\newblock \href {https://doi.org/10.1109/ACCESS.2023.3275764}
  {\path{doi:10.1109/ACCESS.2023.3275764}}.

\bibitem{DBLP:journals/softx/AndaloussiLW23}
A.~A. Andaloussi, D.~L{\"{u}}bke, B.~Weber, Conducting eye-tracking studies on
  large and interactive process models using eyemind, SoftwareX 24 (2023)
  101564.

\bibitem{winter2023comparative}
M.~Winter, C.~Bredemeyer, M.~Reichert, H.~Neumann, R.~Pryss, A comparative
  cross-sectional study on process model comprehension driven by eye tracking
  and electrodermal activity (2023).

\bibitem{DBLP:reference/bdt/Polyvyanyy19}
A.~Polyvyanyy, Business process querying, in: S.~Sakr, A.~Y. Zomaya (Eds.),
  Encyclopedia of Big Data Technologies, Springer, 2019.
\newblock \href {https://doi.org/10.1007/978-3-319-63962-8\_108-1}
  {\path{doi:10.1007/978-3-319-63962-8\_108-1}}.

\bibitem{DBLP:books/sp/22/DelfmannRHCD22}
P.~Delfmann, D.~M. Riehle, S.~H{\"{o}}henberger, C.~Corea, C.~Drodt, The
  diagramed model query language 2.0: Design, implementation, and evaluation,
  in: A.~Polyvyanyy (Ed.), Process Querying Methods, Springer, 2022, pp.
  115--148.

\bibitem{DBLP:books/sp/22/StorrleA22}
H.~St{\"{o}}rrle, V.~Acretoaie, Vm*: {A} family of visual model manipulation
  languages, in: A.~Polyvyanyy (Ed.), Process Querying Methods, Springer, 2022,
  pp. 149--179.

\bibitem{DBLP:books/sp/22/FrancescomarinoT22}
C.~D. Francescomarino, P.~Tonella, The {BPMN} visual query language and process
  querying framework, in: A.~Polyvyanyy (Ed.), Process Querying Methods,
  Springer, 2022, pp. 181--218.

\bibitem{DBLP:books/sp/22/KammererPR22}
K.~Kammerer, R.~Pryss, M.~Reichert, Retrieving, abstracting, and changing
  business process models with {PQL}, in: A.~Polyvyanyy (Ed.), Process Querying
  Methods, Springer, 2022, pp. 219--254.

\bibitem{DBLP:books/sp/22/Polyvyanyy22a}
A.~Polyvyanyy, Process query language, in: A.~Polyvyanyy (Ed.), Process
  Querying Methods, Springer, 2022, pp. 313--341.

\bibitem{DBLP:books/sp/22/ProiettiTS22}
M.~Proietti, F.~Taglino, F.~Smith, Qubpal: Querying business process knowledge,
  in: A.~Polyvyanyy (Ed.), Process Querying Methods, Springer, 2022, pp.
  255--284.

\bibitem{DBLP:journals/eswa/RosaRADMDG11}
M.~L. Rosa, H.~A. Reijers, W.~M.~P. van~der Aalst, R.~M. Dijkman, J.~Mendling,
  M.~Dumas, L.~Garc{\'{\i}}a{-}Ba{\~{n}}uelos, {APROMORE:} an advanced process
  model repository, Expert Syst. Appl. 38~(6) (2011) 7029--7040.

\bibitem{DBLP:conf/abict/LigezaP12}
A.~Ligeza, T.~Potempa, {AI} approach to formal analysis of {BPMN} models:
  Towards a logical model for {BPMN} diagrams, in: M.~Mach{-}Kr{\'{o}}l,
  T.~Pelech{-}Pilichowski (Eds.), {ABICT} 2012, Vol. 257 of Advances in
  Intelligent Systems and Computing, Springer, 2012, pp. 69--88.

\bibitem{DBLP:conf/rcis/CascianiBCM24}
A.~Casciani, M.~L. Bernardi, M.~Cimitile, A.~Marrella, Conversational systems
  for ai-augmented business process management, in: J.~Ara{\'{u}}jo, J.~L.
  de~la Vara, M.~Y. Santos, S.~Assar (Eds.), {RCIS} 2024, Vol. 513 of Lecture
  Notes in Business Information Processing, Springer, 2024, pp. 183--200.

\bibitem{DBLP:conf/bpm/Berti0A23}
A.~Berti, D.~Schuster, W.~M.~P. van~der Aalst, Abstractions, scenarios, and
  prompt definitions for process mining with llms: {A} case study, in: J.~D.
  Weerdt, L.~Pufahl (Eds.), {BPM} 2023 Workshops, Vol. 492 of Lecture Notes in
  Business Information Processing, Springer, 2023, pp. 427--439.

\bibitem{DBLP:journals/corr/abs-2307-12701}
A.~Berti, M.~S. Qafari, Leveraging large language models (llms) for process
  mining (technical report), CoRR abs/2307.12701 (2023).

\bibitem{bernardi2024conversing}
M.~L. Bernardi, A.~Casciani, M.~Cimitile, A.~Marrella, Conversing with business
  process-aware large language models: the bpllm framework (2024).

\bibitem{fahland2024well}
D.~Fahland, F.~Fournier, L.~Limonad, I.~Skarbovsky, A.~J. Swevels, How well can
  large language models explain business processes?, arXiv preprint
  arXiv:2401.12846 (2024).

\bibitem{DBLP:conf/acl/TangKH023}
R.~Tang, D.~Kong, L.~Huang, H.~Xue, Large language models can be lazy learners:
  Analyze shortcuts in in-context learning, in: A.~Rogers, J.~L. Boyd{-}Graber,
  N.~Okazaki (Eds.), Findings of the Association for Computational Linguistics:
  {ACL} 2023, Toronto, Canada, July 9-14, 2023, Association for Computational
  Linguistics, 2023, pp. 4645--4657.

\bibitem{DBLP:journals/corr/abs-2308-07702}
A.~Kong, S.~Zhao, H.~Chen, Q.~Li, Y.~Qin, R.~Sun, X.~Zhou, Better zero-shot
  reasoning with role-play prompting, CoRR abs/2308.07702 (2023).

\bibitem{DBLP:journals/corr/abs-2305-14688}
B.~Xu, A.~Yang, J.~Lin, Q.~Wang, C.~Zhou, Y.~Zhang, Z.~Mao, Expertprompting:
  Instructing large language models to be distinguished experts, CoRR
  abs/2305.14688 (2023).

\bibitem{DBLP:conf/emnlp/JiYXLIF23}
Z.~Ji, T.~Yu, Y.~Xu, N.~Lee, E.~Ishii, P.~Fung, Towards mitigating {LLM}
  hallucination via self reflection, in: H.~Bouamor, J.~Pino, K.~Bali (Eds.),
  Findings of the Association for Computational Linguistics: {EMNLP} 2023,
  Singapore, December 6-10, 2023, Association for Computational Linguistics,
  2023, pp. 1827--1843.

\bibitem{DBLP:journals/corr/abs-2308-02828}
S.~Ouyang, J.~M. Zhang, M.~Harman, M.~Wang, {LLM} is like a box of chocolates:
  the non-determinism of chatgpt in code generation, CoRR abs/2308.02828
  (2023).

\bibitem{DBLP:conf/nips/Wei0SBIXCLZ22}
J.~Wei, X.~Wang, D.~Schuurmans, M.~Bosma, B.~Ichter, F.~Xia, E.~H. Chi, Q.~V.
  Le, D.~Zhou, Chain-of-thought prompting elicits reasoning in large language
  models, in: S.~Koyejo, S.~Mohamed, A.~Agarwal, D.~Belgrave, K.~Cho, A.~Oh
  (Eds.), NeurIPS 2022, 2022.

\bibitem{DBLP:conf/icmla/KaziK23}
N.~Kazi, I.~Kahanda, Enhancing transfer learning of llms through fine- tuning
  on task - related corpora for automated short-answer grading, in:
  International Conference on Machine Learning and Applications, {ICMLA} 2023,
  Jacksonville, FL, USA, December 15-17, 2023, {IEEE}, 2023, pp. 1687--1691.

\bibitem{DBLP:conf/esws/MartinoIT23}
A.~Martino, M.~Iannelli, C.~Truong, Knowledge injection to counter large
  language model {(LLM)} hallucination, in: C.~Pesquita, H.~Skaf{-}Molli,
  V.~Efthymiou, S.~Kirrane, A.~Ngonga, D.~Collarana, R.~Cerqueira, M.~Alam,
  C.~Trojahn, S.~Hertling (Eds.), {ESWC} 2023 Satellite Events, Vol. 13998 of
  Lecture Notes in Computer Science, Springer, 2023, pp. 182--185.

\bibitem{DBLP:conf/nips/BrownMRSKDNSSAA20}
T.~B. Brown, B.~Mann, N.~Ryder, M.~S. et~al., Language models are few-shot
  learners, in: H.~Larochelle, M.~Ranzato, R.~Hadsell, M.~Balcan, H.~Lin
  (Eds.), NeurIPS 2020, 2020.

\bibitem{DBLP:journals/corr/abs-2305-16807}
D.~Miyake, A.~Iohara, Y.~Saito, T.~Tanaka, Negative-prompt inversion: Fast
  image inversion for editing with text-guided diffusion models, CoRR
  abs/2305.16807 (2023).

\bibitem{munoz-gama2019}
J.~Munoz-Gama, R.~de~la Fuente, M.~Sepúlveda, R.~Fuentes, Conformance checking
  challenge 2019 (2019).

\bibitem{DBLP:conf/bpmds/BertiKHL024}
A.~Berti, H.~Kourani, H.~H{\"{a}}fke, C.~Li, D.~Schuster, Evaluating large
  language models in process mining: Capabilities, benchmarks, and evaluation
  strategies, in: H.~van~der Aa, D.~Bork, R.~Schmidt, A.~Sturm (Eds.),
  Enterprise, Business-Process and Information Systems Modeling - 25th
  International Conference, {BPMDS} 2024, and 29th International Conference,
  {EMMSAD} 2024, Limassol, Cyprus, June 3-4, 2024, Proceedings, Vol. 511 of
  Lecture Notes in Business Information Processing, Springer, 2024, pp. 13--21.
\newblock \href {https://doi.org/10.1007/978-3-031-61007-3\_2}
  {\path{doi:10.1007/978-3-031-61007-3\_2}}.

\bibitem{DBLP:journals/corr/abs-2310-06271}
Z.~Ji, T.~Yu, Y.~Xu, N.~Lee, E.~Ishii, P.~Fung, Towards mitigating
  hallucination in large language models via self-reflection, CoRR
  abs/2310.06271 (2023).

\bibitem{DBLP:journals/iando/SchultzeA11}
U.~Schultze, M.~Avital, Designing interviews to generate rich data for
  information systems research, Inf. Organ. 21~(1) (2011) 1--16.
\newblock \href {https://doi.org/10.1016/J.INFOANDORG.2010.11.001}
  {\path{doi:10.1016/J.INFOANDORG.2010.11.001}}.

\bibitem{DBLP:conf/fat/SokolF20}
K.~Sokol, P.~A. Flach, Explainability fact sheets: a framework for systematic
  assessment of explainable approaches, in: M.~Hildebrandt, C.~Castillo, L.~E.
  Celis, S.~Ruggieri, L.~Taylor, G.~Zanfir{-}Fortuna (Eds.), FAT* '20, {ACM},
  2020, pp. 56--67.

\bibitem{DBLP:conf/nips/LewisPPPKGKLYR020}
P.~S. H.~L. et~al., Retrieval-augmented generation for knowledge-intensive
  {NLP} tasks, in: H.~Larochelle, M.~Ranzato, R.~Hadsell, M.~Balcan, H.~Lin
  (Eds.), Advances in Neural Information Processing Systems 33: Annual
  Conference on Neural Information Processing Systems 2020, NeurIPS 2020,
  December 6-12, 2020, virtual, 2020.

\bibitem{DBLP:conf/bpmds/KouraniB0A24}
H.~Kourani, A.~Berti, D.~Schuster, W.~M.~P. van~der Aalst, Process modeling
  with large language models, in: H.~van~der Aa, D.~Bork, R.~Schmidt, A.~Sturm
  (Eds.), Enterprise, Business-Process and Information Systems Modeling - 25th
  International Conference, {BPMDS} 2024, and 29th International Conference,
  {EMMSAD} 2024, Limassol, Cyprus, June 3-4, 2024, Proceedings, Vol. 511 of
  Lecture Notes in Business Information Processing, Springer, 2024, pp.
  229--244.
\newblock \href {https://doi.org/10.1007/978-3-031-61007-3\_18}
  {\path{doi:10.1007/978-3-031-61007-3\_18}}.

\bibitem{DBLP:journals/corr/abs-2403-04327}
H.~Kourani, A.~Berti, D.~Schuster, W.~M.~P. van~der Aalst, {ProMoAI}: Process
  modeling with generative {AI}, CoRR abs/2403.04327 (2024).
\newblock \href {http://arxiv.org/abs/2403.04327} {\path{arXiv:2403.04327}},
  \href {https://doi.org/10.48550/ARXIV.2403.04327}
  {\path{doi:10.48550/ARXIV.2403.04327}}.

\bibitem{DBLP:books/sp/Aalst16}
W.~M.~P. van~der Aalst,
  \href{https://doi.org/10.1007/978-3-662-49851-4}{Process Mining - Data
  Science in Action, Second Edition}, Springer, 2016.
\newblock \href {https://doi.org/10.1007/978-3-662-49851-4}
  {\path{doi:10.1007/978-3-662-49851-4}}.
\newline\urlprefix\url{https://doi.org/10.1007/978-3-662-49851-4}

\end{thebibliography}

\newpage\clearpage

\appendix






\end{document}
\endinput